\documentclass[12pt,preprint]{aastex}

\usepackage{graphicx}
\usepackage{amsmath}
\usepackage{amssymb}
\usepackage{latexsym}
\usepackage{epsfig}
\usepackage{natbib}

\newcommand{\ebold}{{\boldsymbol{e}}}

\newcommand{\ibold}{{\boldsymbol{i}}}
\newcommand{\ubold}{{\boldsymbol{v}}}
\newcommand{\xbold}{{\boldsymbol{x}}}
\newcommand{\edbold}{{\ebold^{d}}}
\newcommand{\eboldone}{{\ebold^{d_1}}}
\newcommand{\eboldtwo}{{\ebold^{d_2}}}
\newcommand{\Dim}{{\mathbf{D}}}
\newcommand{\half}{\frac{1}{2}}
\newcommand{\fourth}{\frac{1}{4}}

\begin{document}
\title{Constrained-Transport Magnetohydrodynamics with Adaptive-Mesh-Refinement in {\tt CHARM}}

\author{Francesco Miniati}
\affil{Physics Department, Wolfgang-Pauli-Strasse 27,
ETH-Z\"urich, CH-8093, Z\"urich, Switzerland; fm@phys.ethz.ch}
\author{Daniel F. Martin}
\affil{Lawrence Berkeley National Laboratory, 1 Cyclotron Rd, Berkeley, CA 94720, USA; DFMartin@lbl.gov}


\begin{abstract}
  We present the implementation of a three-dimensional, second order
  accurate Godunov-type algorithm for magneto-hydrodynamic (MHD), in
  the adaptive-mesh-refinement (AMR) cosmological code {\tt CHARM}.
  The algorithm is based on the full 12-solve spatially unsplit
  Corner-Transport-Upwind (CTU) scheme.  The fluid quantities are
  cell-centered and are updated using the Piecewise-Parabolic-Method
  (PPM), while the magnetic field variables are face-centered and are
  evolved through application of the Stokes theorem on cell edges via a
  Constrained-Transport (CT) method. 
  The multidimensional MHD source terms required in the
  predictor step for high-order accuracy are applied in a simplified
  form which reduces their complexity in three dimensions
  without loss of accuracy or robustness.  The algorithm is
  implemented on an AMR framework which requires specific
  synchronization steps across refinement levels.  These include
  face-centered restriction and prolongation operations and a {\it
    reflux-curl} operation, which maintains a solenoidal magnetic field
  across refinement boundaries.  The code is tested against a large
  suite of test problems, including convergence tests in smooth flows,
  shock-tube tests, classical two- and three-dimensional MHD tests,
  a three-dimensional shock-cloud interaction problem and the
  formation of a cluster of galaxies in a fully cosmological
  context. The magnetic field divergence is shown to remain
  negligible throughout.

\end{abstract}
\keywords{cosmology: theory --- methods: numerical --- magnetohydrodynamics: MHD}

\section{Introduction} \label{intro:sec}

Magnetic fields are a common feature of cosmic plasmas, from
the interplanetary medium and the atmospheres of stars, to the
interstellar medium of galaxies and the baryonic gas in the largest
structures of the universe such as clusters and voids of
galaxies~\citep[e.g.,][]{zeruso83,bernetetal08,clkrbo00,neronovandvovk10}.

Their origin is discussed in several papers and different processes
are likely responsible for the magnetization of different
environments~\citep[see, e.g.][and references
therein]{Miniati2011ApJ...729...73M}. In general weak rotational
electrostatic fields are required, which are normally suppressed by
the high conductivity of astrophysical plasmas, but which can
nevertheless arise under special conditions.  The magnetic field can
then evolve considerably and amplification of an initially weak seed
by many orders of magnitude is plausible, particularly when the flow
is highly turbulent~\citep{zeruso83}.  Magnetic fields affect the
dynamics of a system directly through the Lorentz force and indirectly
through their impact on the plasma microscopic properties, e.g. the
thermal conductivity or electric resistivity~\citep{Spitzer1965pfig.book.....S},
or the transport of high energy particles~\citep{Schlickeiser2002}, both
with conspicuous macroscopic effects.
Thus magnetic fields are a crucial
component of astrophysical plasmas although perhaps due to the
complexity they introduce, progress in characterizing their role has
been relatively slow, particularly in certain areas of astrophysics.
Due to such complexities, particularly during highly nonlinear
regimes, accurate and efficient numerical methods are
valuable for studying the evolution of magnetized systems.

A simplified description of a magnetized fluid is provided by the equations 
of ideal magneto-hydrodynamics (MHD). This approximation is 
valid when the following hierarchy of scales
is satisfied: $\ell_{mfp}\ll \lambda \ll L$, where $\ell_{mfp}$ is the
mean-free-path of the fluid particles, $\lambda$ is the characteristic
scale of the problem of interest and $L$ is the size of the 
system~\citep{ginzburg79}.
The $mfp$ can be suppressed considerably in the direction
perpendicular to the magnetic field lines, but in parallel directions one
relies typically on collisions to guarantee the fluid approximation.
In reality microscopic plasma processes or a tangled
component of the magnetic field can also provide the required
viscosity. Notably the MHD approximation prevents the occurrence of kinetic
processes which, however, can sometimes by represented by source 
terms on the RHS of the MHD equations.

When dissipative terms can be neglected, the MHD equations are in
ideal form and read~\citep{lali8}
\begin{eqnarray}
\label{rho:eq}
\frac{\partial\rho}{\partial t} + \frac{\partial \rho u_j}{\partial x_j}=0,\\  
\label{mom:eq}
\frac{\partial\rho u_i}{\partial t} + \frac{\partial}{\partial x_j} 
\left( \rho u_iu_j+p\delta_{ij}-B_jB_i\right)=0,\\
\label{ene:eq}
\frac{\partial\rho e}{\partial t} + \frac{\partial}{\partial x_j} 
\left[u_j\left( \rho e+p\right)-B_jB_iu_i\right] = 0.
\end{eqnarray}
Here $\rho$ is the gas density, $u_i$ and $B_i$ the components of the
velocity and magnetic field vectors, respectively, $p=p_g+\half
B^2$ is the total sum of the gas and magnetic pressures, $e=\half
u^2+ e_{th}+\half \frac{B^2}{\rho}$ is the total specific energy
density, $\delta_{ij}$ Kronecker's delta and summation over repeated
indices is assumed.  The thermal energy is related to the pressure
through a $\gamma$-law equation of state, $e_{th}=p_g/\rho
(\gamma-1).$ The magnetic field evolution is described by Faraday's
equation, with the electric field given by Ohm's law. In the
limit of negligible resistivity the only electric fields are those
induced by motions of the magnetized fluid and the induction equation 
reads~\citep{lali8}
\begin{equation} \label{faraday:eq}
\frac{\partial B_i}{\partial t} 
= -\varepsilon_{ijk} \frac{\partial E_k}{\partial x_j}
= -\frac{\partial}{\partial x_j} \left(u_jB_i-B_ju_i\right),
\end{equation}
where $\varepsilon_{ijk}$ is the fully antisymmetric tensor of rank 3
and $\varepsilon_{012}=1$.

The resistive terms neglected in Eq.
(\ref{rho:eq})--(\ref{faraday:eq}), which are responsible for the diffusion of the
magnetic field,  can be readily recovered~\citep{Samtaney2005JPhCS..16...40S}, although for most purposes their neglect is safe in astrophysical plasmas.

There are various approaches to solve numerically the equations of MHD.
The one we follow in this paper is based on the extension to MHD of
conservative methods for hyperbolic systems of equations, particularly Godunov's
methods for hydrodynamics. This approach is met with two difficulties, however.
First, care much be taken
because the system of MHD equations is not strictly hyperbolic. 
This can be dealt with
by ``renormalizing'' the eigenvalues and eigenvectors of the system~\citep{BrioWu1988JCoPh..75..400B}.  Second, and most importantly the solenoidal constraint, 
${\bf \nabla\cdot B}=0$, must be enforced or else the solution will contain artifacts~\citep{BrackbillBarnes1980JCoPh..35..426B}.
Unfortunately numerical 
schemes designed for pure hydrodynamics do not fulfill such requirement
and the above constraint must be enforced separately. 
Different ways to do so have been proposed.
In one approach the magnetic field is defined together with all other fluid 
variables at cell centers and the non-solenoidal component is removed through
a Hodge-Helmholtz projection 
method, typically once per time-step~\citep{BrackbillBarnes1980JCoPh..35..426B,ZacharyMalagoliColella1994,RYUETAL1995ApJ...452..785R}.
In a variation of this approach the projection operation is 
performed on the magnetic field variables extrapolated to the
cell faces which are used to define the MHD fluxes~\citep{Crockett2005JCoPh.203..422C}. 
It is argued in~\citet{Crockett2005JCoPh.203..422C} 
that this type of projection is mathematically more consistent
with the solenoidal requirement because it is the fluxes defined on
the cell faces that get differentiated to compute the flux updates.
In a second approach, the MHD equations are cast in a special
8-wave non-conservative formulation. This allows for the non-solenoidal
component of the magnetic field to be explicitly tracked and suitably
damped as it is advected by the
flow~\citep{Powell1999JCoPh.154..284P,Dedner2002JCoPh.175..645D}.

Finally in the Constrained Transport (CT) approach, the discretization strategy
originally proposed by~\cite{Yee1966} in the context of Maxwell
equations is used, in which the magnetic field is defined at face centers,
the electric field used to update the latter is defined at cell edges,
and the other fluid variables are defined at cell centers as in ordinary
hydrodynamics~\citep{EvansHawley1988ApJ...332..659E,DaiWoodward1998ApJ...494..317D,rmjf98,
  BalsaraSpicer1999JCoPh.149..270B,Toth2000JCoPh.161..605T,LondrillodelZanna2004JCoPh.195...17L,Fromang2006A&A...457..371F,Cunningham2009ApJS..182..519C}.
In this approach the rate of change of the magnetic flux at cell faces
is given by the circulation of the electric field along the cell edges
which define the boundary of the corresponding face. Thus the
solenoidal character of the magnetic field is ensured by Stokes'
theorem down to machine precision.
Recently~\citet{GardinerStone2005JCoPh.205..509G,GardinerStone2008JCoPh.227.4123G}
have developed an unsplit version of the CT algorithm, extending to
the case of MHD the Corner Transport Upwind (CTU) method for
directionally unsplit hydrodynamics proposed in~\citet{Colella1990JCoPh..87..171C} and
\citet{saltzman94}.  The use of a directionally unsplit algorithm has
proven quite attractive, particularly because it appears to be better
suited for modeling turbulent flows (Bell et al) and in the presence of
source terms~\citep{Leveque1998cmaf.conf....1L}. Most importantly, due to the solenoidal constraint, the
MHD equations contain intrinsically multidimensional terms which
require a directionally unsplit formulation if one is to achieve high
order accuracy in multidimensional
problems~\citep{GardinerStone2005JCoPh.205..509G}.  Thus the extension of CTU
to MHD has also attracted considerable interest in the astrophysical
community~\cite[e.g.][]{Teyssier2006JCoPh.218...44T,Fromang2006A&A...457..371F,
  LeeDeane2009JCoPh.228..952L,MignoneTzeferacos2010JCoPh.229.2117M}.

In this paper we describe the implementation of a version of the CTU +
CT scheme that closely resembles the one
of~\citet{GardinerStone2008JCoPh.227.4123G}
and~\citet{Stone2008ApJS..178..137S}.  Our scheme differs from theirs, however,
in two respects. First we have chosen to use the full
12-solve CTU scheme instead of the simpler 6-solve scheme.  The reason
for this choice is due to the larger CFL number that the full CTU
scheme can afford (CFL=1), as compared to the simplified version
(CFL=0.5). As indicated by~\citet{GardinerStone2008JCoPh.227.4123G}
the computational
cost of the two versions of the CTU scheme is roughly the same for
pure MHD calculations, because the factor of two fewer Riemann solvers
comes with twice as many steps to achieve the same solution time.
However, for multi-physics 
applications that we have in mind other expensive solvers are
executed each time step, whose cost grows with the number of
time-steps.
Note that~\citet{Teyssier2006JCoPh.218...44T} use also a
simpler version of the CTU scheme, although in their case a slightly
less restrictive condition on the time-step is nominally allowed, i.e.
CFL$\le0.7$. On the other hand, Lee \& Deane (2009) have introduced a
different approach for computing the transverse flux updates of the
fluid variables in their directionally unsplit method, which relies on
characteristic tracing alone and does not require intermediate Riemann
solvers (except for the magnetic field intermediate updates).  For
this reason and since the stability constraint only requires CFL$\le1$
this approach can potentially be quite efficient.

Secondly, we take into account the multidimensional corrections
required to balance the ${\bf \nabla\cdot B}$ terms in a form that is 
simpler than originally proposed by~\citet{GardinerStone2008JCoPh.227.4123G}, 
and analogous to~\citet{Crockett2005JCoPh.203..422C}. Our tests suggest that
the accuracy and robustness of the algorithm are not affected by this simplification.

Finally, using in particular the ideas in~\cite{bergercolella89} for
adaptive-mesh-refinement (AMR) and~\citet{Balsara2001JCoPh.174..614B} for the
divergence-free coarse-fine interpolation of the magnetic field in
newly refined grid patches, we have implemented an AMR version of our
CTU + CT MHD scheme.  The code in which the implementation is carried
out is {\tt CHARM}~\citep{mico07b}.  It includes various
other physical modules, namely self-gravity, collision-less
dark matter particles and
cosmic expansion for cosmological applications, radiative cooling
~\citep{mico07a}, cosmic-rays~\citep{min01,min07} and dust
particles~\citep{miniati10}.

The remainder of this paper is organized as follows.  The algorithm is
discussed in detail in Sec.~\ref{numsche::se}. 
In Sec~\ref{tests:se} we present results for an extensive set of problems
that test the accuracy and robustness of the code.  These tests
include a convergence study in smooth flows, a suite of Riemann
problems in one and two dimensions, the Orszag-Tang Vortex as well as
the rotor problem, carried out on a uniform grid.
Finally, extension to AMR and
to the case of cosmological applications are described in
Sec.~\ref{amr::se} and~\ref{cosmol::se}, respectively.  We have tested
these extensions with a problem involving the interaction of a magnetized 
interstellar cloud with a strong shock, and the formation of a galaxy cluster
in a fully cosmological simulation. The paper closes with a brief summary in
 Sec.~\ref{summary:se}.
%

%
%
\section{Numerical Scheme} \label{numsche::se}

In this section we first provide an overall description of the full CTU + CT
algorithm and, following that, we discuss the implementation
details.  We begin by a description of the space discretization, data
structure and notation.

\subsection{Preliminaries} \label{prel:se}
\subsubsection{Discretization, Variables and Operators} \label{dvo:se}

\begin{figure*}[t]
\epsscale{1}\plottwo{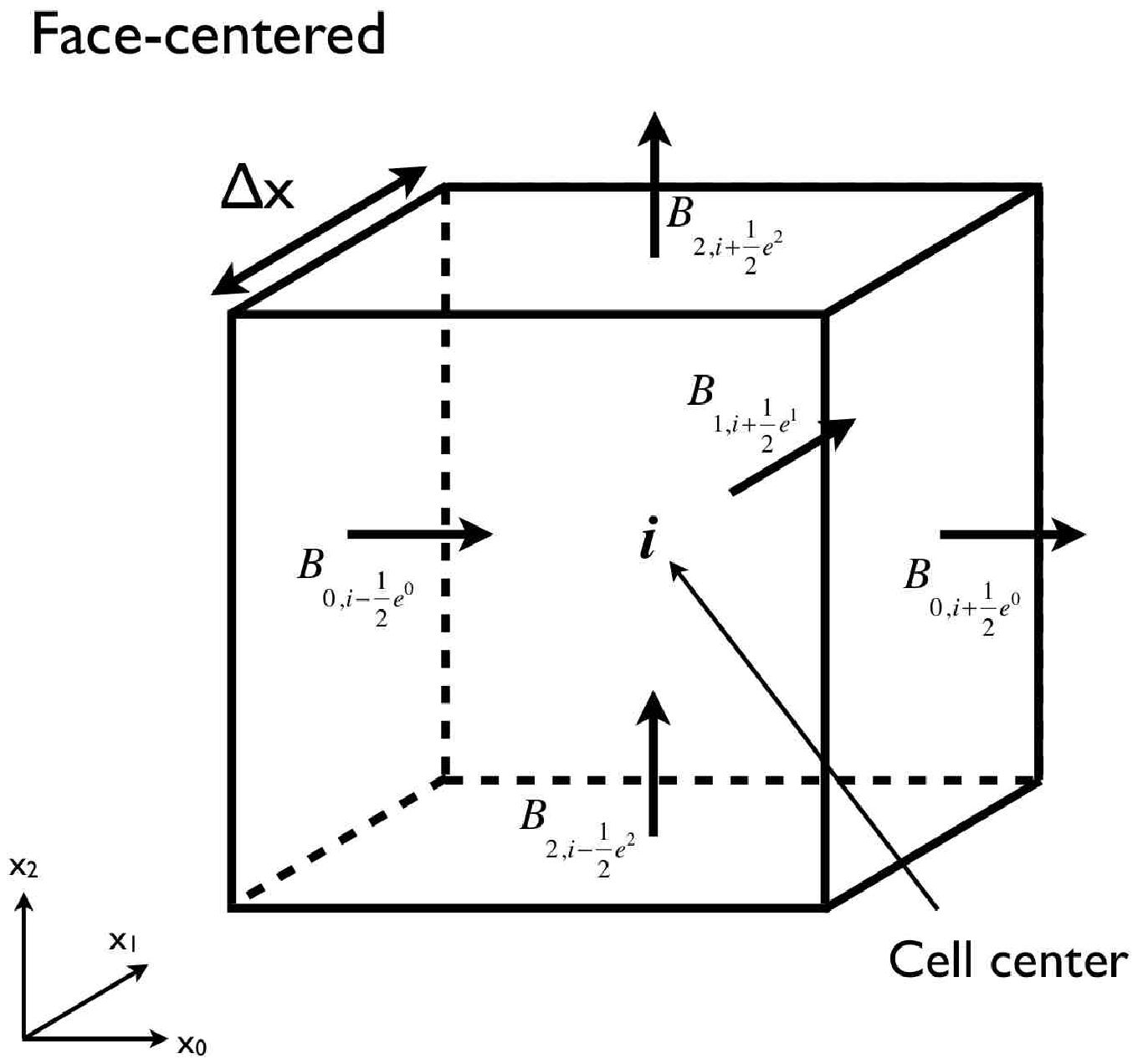}{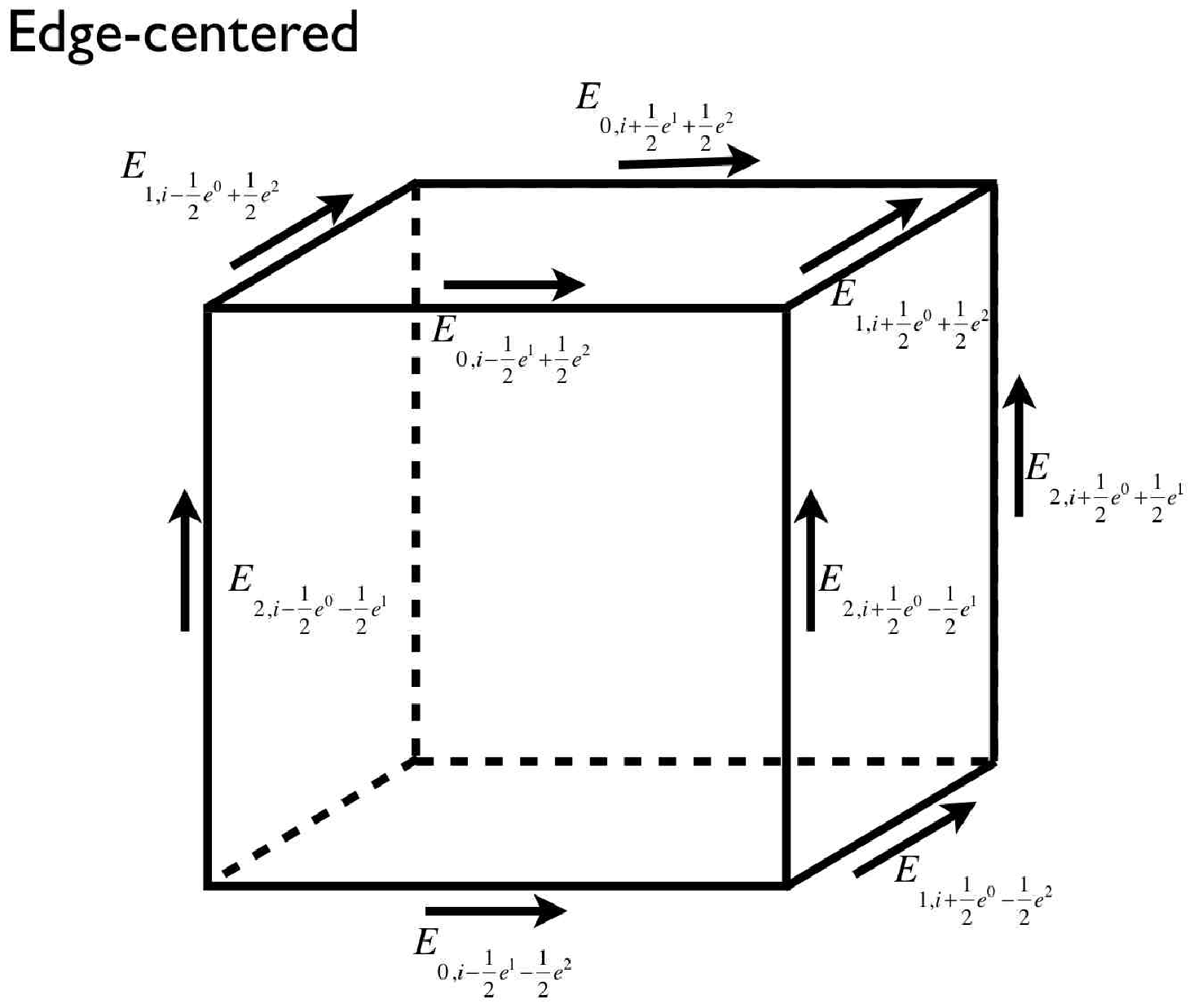} \\
\caption{Control volume and discrete representation of physical quantities.}
\label{cell:fig}
\end{figure*}

The algorithmic operations are carried out on a discrete
representation of the continuous $\Dim$-dimensional space given by the
cubic lattice, ${\ibold}\equiv(i_0, ... , i_{\Dim - 1}) \in
\mathbb{Z}^\Dim$.  The computational domain, referred to as a
grid $\Gamma$, is a bound subset of $\mathbb{Z}^\Dim$ and
provides a finite-volume discretization of the continuous space
into a collection of control volumes, faces and edges. Each control volume is
identified by an index ${\ibold}\equiv(i_0, ... , i_{\Dim - 1}) \in
\Gamma$ and corresponds to a region of space,
\begin{equation}
V_{\ibold} = [\ibold h, (\ibold + \ubold)h],
\end{equation}
where $h$ is the mesh spacing, and $\ubold \equiv (1,...,1)$ is the
vector whose components are all equal to one.  The
face-centered discretization of space based on the same control
volumes is: $\Gamma_{\rm f}^{\edbold} = \{ \ibold \pm \half \edbold :
\ibold \in \Gamma\} $, where $\edbold$ is the unit vector in the $d$
direction.  $\Gamma_{\rm f}^{\edbold}$ 
indexes the cell faces of $\Gamma$ normal to $\edbold$ representing the areas

\begin{equation}
  A_{\ibold + \half \edbold} =  [(\ibold + \edbold )h, (\ibold + \ubold)h], \quad\ibold + \half \edbold \in \Gamma_{\rm f}^{\edbold}.
\end{equation}
Finally, the edge-centered discretization of space is:
$\Gamma_{\rm e}^{\edbold} = \{ \ibold \pm \half \eboldone \pm\half
\eboldtwo : \ibold \in \Gamma,\,d\ne d_1\ne d_2\} $.  $\Gamma_{\rm e}^{\edbold}$ 
indexes the edges of the cells in $\Gamma$ aligned with $\edbold$ 
representing the lengths
\begin{equation}
L_{\ibold + \half (\eboldone+\eboldtwo)} =  [\ibold h+\half (\eboldone+\eboldtwo), (\ibold + \ubold)h],\quad \ibold+ \half (\eboldone+\eboldtwo) \in \Gamma_{\rm e}^{\edbold}.
\end{equation}
Fig.~\ref{cell:fig} illustrates a control volume with the various 
types of discretization.

Given the above discretization, we define
cell-centered discrete variables on $\Gamma$ as
\begin{gather*}
\phi : \Gamma \rightarrow {\mathbb{R}}^m,
\end{gather*}
and denote by $\phi_\ibold \in {\mathbb{R}}^m$ the value of $\phi$ at 
cell $\ibold \in \Gamma$.  Similarly we define face-centered vector fields 
on $\Gamma_{\rm f}^{\edbold}$ as
\begin{gather*}
{\vec F} = (F_0,\dots,F_{\Dim-1}) \hbox{ , } F_d : \Gamma_{\rm f}^{\ebold^d}
\rightarrow {\mathbb{R}}^m,
\end{gather*}
and denote by $F_{d,\ibold + \half \edbold} \in {\mathbb{R}}^m$ the
value of $F_d$ at $\ibold + \half \edbold \in \Gamma_{\rm
  f}^{\edbold}$, and also define edge-centered vector fields on
$\Gamma_{\rm e}^{\edbold}$ as
\begin{gather*}
{\vec E} = (E_0,\dots, E_{\Dim-1}) \hbox{ , } E_d : \Gamma_{\rm e}^{\ebold^d}
\rightarrow {\mathbb{R}}^m
\end{gather*}
and denote by $E_{d,\ibold + \half (\eboldone+\eboldtwo)} \in {\mathbb{R}}^m$
the value of $E_d$
at $ \ibold+ \half (\eboldone+\eboldtwo) \in \Gamma_{\rm e}^{\edbold}.$
Finally, for face-centered fields we introduce the discretized divergence 
operator
\begin{equation} \label{div_def:eq}
\left(\vec D\cdot \vec{F}\right)_\ibold = \frac{1}{h} \sum^{\Dim-1}_{d=0} \left(F_{d,\ibold
+\half \ebold^d} - F_{d,\ibold - \half \ebold^d}\right), \ibold \in
\Gamma,
\end{equation}
and for edge-centered fields the discretized curl operator
\begin{equation} \label{curl_def:eq}
\left(\vec D\times \vec{E}\right)_{d,\ibold+\half\edbold} = \frac{1}{h} \sum_{d_1, d_2}
 \varepsilon_{dd_1d_2} \left(E_{d_1,\ibold+\half\ebold^d+\half\eboldtwo} - E_{d_1,\ibold +\half \ebold^d-\half\eboldtwo}\right), 
 \quad\ibold + \half \edbold \in \Gamma_{\rm f}^{\edbold}.
\end{equation}

Time is also discretized into a number of finite intervals of
variable size. In particular, we write $t^{n+1}=t^n+\Delta t^n$, where
$t$ indicates the solution time, $n$ indicates the integration step,
and $\Delta t^n$ the time-step interval.

\subsubsection{Time Integration} \label{timeint:se}

The coupled system of equations (\ref{rho:eq})-(\ref{faraday:eq}),
with the allowance for a non-zero source term $S(U)$,
can be cast in the following compact form
\begin{equation}\label{cons_sys:eq}
\frac{\partial U}{\partial t} + {\bf\nabla\cdot F} = S,
\end{equation}
where the conservative variables, $U$, and the associated fluxes 
along the direction $d$, $F_d$, are defined as
\begin{equation} \label{cons_uf:eq}
U=
\begin{pmatrix}
\rho \\
\rho u_0 \\
\vdots \\
\rho u_{\Dim-1} \\
\rho e  \\
B_0 \\
\vdots \\
B_{\Dim-1}
\end{pmatrix}
, \quad
F_d(U)=
\begin{pmatrix}
\rho u_d \\
\rho u_0 u_d + p \,\delta_{d0} -B_0B_d \\
\vdots \\
\rho u_{\Dim-1} u_d + p\,\delta_{d\Dim-1} -B_{\Dim-1}B_d  \\
u_d\left( \rho e+p\right)-B_d{\bf B\cdot u} \\
u_0B_d-B_0u_d \\
\vdots \\
u_{\Dim-1}B_d-B_{\Dim-1}u_d
\end{pmatrix}.
\end{equation}
Following Godunov's approach and its
higher order extensions, we can conveniently formulate a
numerical integration scheme based on the conservative properties of
the system (\ref{cons_sys:eq}).  In this approach one follows the
evolution of cell-centered volume-averaged conservative variables,
defined as
\begin{equation}\label{vau:eq}
U^n_\ibold= \frac{1}{V_\ibold} \int_{V_\ibold} U(t^n,\xbold)\,dV.
\end{equation}
The evolution equation is obtained upon suitable manipulation
of the original continuous Eq. (\ref{cons_sys:eq}), and 
reads~\citep{Leveque1998cmaf.conf....1L}
\begin{equation}\label{cons_fv:eq}
U_\ibold^{n+1} = U_\ibold^n- 
\frac{\Delta t}{\Delta x}\, \sum^{\Dim-1}_{d=0} 
\left(F^{n+\half}_{d,\ibold +\half \ebold^d} - 
F^{n+\half}_{d,\ibold - \half \ebold^d}\right) + S^{n+\half},
\end{equation}
where the face-centered time-averaged fluxes along the direction $d$
are defined as
\begin{equation}\label{taf:eq}
F^{n+\half}_{d,\ibold +\half \ebold^d} = \frac{1}{\Delta t A_{\ibold +\half \ebold^d}} \int_{t^n}^{t^{n+1}}dt\int_{A_{\ibold +\half \ebold^d}} F_d(t,\xbold) \,dA.
\end{equation}
If the source term is non-stiff,
we can obtain a second order estimate for it by the simple time-average, 
$S^{n+\half}\simeq\half(S^n+S^{n+1})$.

Note that given the flux along a direction $d_1$, $F_{d_1,\ibold +\half\eboldone}$,
the component corresponding to the magnetic field along the direction $d_2\ne d_1$,
defines a face centered electric field along $d$ according to
\begin{equation}\label{fcef:eq}
E_{d,\ibold+\half\eboldone}=-\varepsilon_{dd_1d_2}\,F_{d_1,\ibold +\half\eboldone}
^{B_{d_2}}(U),
\end{equation}
where we use subscripts to indicate directions and centering, and superscripts 
for components.
Indeed, Godunov's scheme also updates in time
the cell-centered magnetic fields variables. As already pointed out,
however, the updated magnetic field in general does not remain solenoidal. 
Therefore, we adopt instead a CT discretization strategy in which 
the primary description of the magnetic field we will be
using face-centered area-averaged variables defined as
\begin{equation}\label{fab:eq}
B^n_{d,\ibold +\half \ebold^d} = \frac{1}{A_{\ibold +\half \ebold^d}} 
\int_{A_{\ibold +\half \ebold^d}} B_d(t^n,\xbold) \,dA.
\end{equation}
Note that an estimate of the cell-centered magnetic field variables is
still needed in order to construct the fluxes in (\ref{taf:eq}). We
will return to this point shortly.  The evolution equation for the
face-centered magnetic field variables is obtained again from a
manipulation of Faraday's law and reads
\begin{alignat}{2}
\notag 
B_{d,\ibold +\half \ebold^d}^{n+1} = B_{d,\ibold +\half \ebold^d}^n 
-&\varepsilon_{dd_1d_2} \frac{\Delta t}{\Delta x}
 \left[ \left(
E^{n+\half}_{d_1,\ibold+\half\ebold^d+\half\eboldtwo} - 
E^{n+\half}_{d_1,\ibold +\half \ebold^d-\half\eboldtwo}\right)\right. 
\\ & \left.  -\left(E^{n+\half}_{d_2,\ibold+\half\ebold^d+\half\eboldone} - 
E^{n+\half}_{d_2,\ibold +\half \ebold^d-\half\eboldone}\right)\right] 
\notag \\
+& S_{B_d,\ibold +\half \ebold^d}^{n+\half}
\label{faraday_fa:eq}
\end{alignat}
with $d \neq d_1 \neq d_2,0 \le d, d_1, d_2 < \Dim$.
The edge-centered time-averaged electric field is formally
defined as
\begin{equation}\label{tae:eq}
E^{n+\half}_{d,\ibold +\half\eboldone+\half\eboldtwo} = 
\frac{1}{\Delta t\,L_{\ibold + \half (\eboldone+\eboldtwo)}} 
\int_{t^n}^{t^{n+1}}dt\int_{L_{\ibold + \half (\eboldone+\eboldtwo)}} E_d(t,\xbold) \,dL.
\end{equation}
and, as above, the time-centered estimate of the non-stiff source term
for the face-centered magnetic field components is obtained by simple
arithmetic averaging.
The cell-centered magnetic field variables are then defined in terms
of the primary face-centered values using the following second-order
accurate reconstruction
scheme~\citep{rmjf98,Balsara2001JCoPh.174..614B,GardinerStone2005JCoPh.205..509G}
\begin{equation}\label{bcc:eq}
B_{d,\ibold} = \half \left(
B_{d,\ibold -\half \ebold^d}+B_{d,\ibold +\half \ebold^d}
\right).
\end{equation}
An important part of CT schemes concerns the calculation of the
time-averaged, edge-centered electric fields entering the time
update~(\ref{faraday_fa:eq}). In general one employs some type of
bilinear interpolation.~\cite{DaiWoodward1998ApJ...494..317D}
interpolate in space and time the cell-centered velocity and magnetic
field at time $t^n$ with the solution from the Godunov scheme at time
$t^{n+1}$ to obtain time- and corner-centered electric fields.  On
the other hand,~\cite{rmjf98} and~\cite{Balsara2001JCoPh.174..614B}
take advantage of Eq.~(\ref{fcef:eq}) and interpolate the
face-centered variables that allow reconstruction of electric fields
at cell edges.  The interpolation scheme proposed in~\cite{rmjf98} has
the property that for plane-parallel configurations their
multidimensional scheme reduces to the one-dimensional scheme.  This
same property is shared by the upwind scheme proposed
in~\cite{GardinerStone2005JCoPh.205..509G} which is adopted here and
is described in more detail in Sec.~\ref{ecef:sec}. This
property is important because it guarantees self-consistency between
(a) the electric fields used for the time-update of
face-centered magnetic variables, (b) the MHD fluxes used for the time-update of
the cell-centered variables and (c) the
synchronization step of the magnetic variables given in
Eq.~(\ref{bcc:eq})~\citep{GardinerStone2005JCoPh.205..509G}.

Note that by applying the divergence operator in
Eq. (\ref{div_def:eq}) to the face-centered magnetic field
evolved according to Eq. (\ref{faraday_fa:eq}), one finds that the
divergence of the field does not change in time. So the magnetic field
remains solenoidal if initially so. This is the property of the CT
scheme.

The accuracy and stability of the numerical solution depend
principally on how the time-averaged fluxes and electric fields
entering Eq. (\ref{cons_fv:eq}) and (\ref{faraday_fa:eq}),
respectively, are computed.  In the following sections we shall
describe the algorithmic details characterizing such calculation.

\subsection{Algorithm}  \label{algo:se}

The algorithm described in this section computes second-order accurate face
centered fluxes and edge-centered electric fields required for the
update of the cell-centered fluid variables and face-centered magnetic
field variables in Eq.(\ref{cons_fv:eq}) and (\ref{faraday_fa:eq}),
respectively. It uses a combination of the CTU
algorithm~\citep{Colella1990JCoPh..87..171C,saltzman94}, 
and the CT scheme, as in~\citet{GardinerStone2008JCoPh.227.4123G}.
Unlike these authors, though, for the reasons indicated
in the Introduction we will employ the full CTU scheme with 12-solve
per cells (in 3-dimensions).

Assuming the solution at time $t^n$, $U^n_\ibold$ to be known, the outline
of the algorithm is as follows:

\begin{enumerate}
\item transform from conservative to primitive variables,
 $U^n_\ibold\leftarrow W^n_\ibold$, and synchronize cell-centered and
  face-centered magnetic field values
\item compute limited slopes along the coordinate directions, $\delta
  W^n_{\ibold, d}$.
\item do characteristic tracing to extrapolate in space and time
  primitive variables from cell centers to cell faces, $W_{\ibold,
    \pm, d}$. Also include effects of source term here.
\item convert back to conservative variables,
$W_{\ibold, \pm, d}\leftarrow U_{\ibold, \pm, d}$.
\item apply corrections to $U_{\ibold, \pm, d}$ due to transverse
  gradients and obtain multidimensionally correct time-averaged fluxes
  $F_{\ibold+\half\edbold}^{n+\half}$ and electric fields
  $E_{\ibold+\half\eboldone+\half\eboldtwo}^{n+\half}$.
\item update primary variables, $U_\ibold^n\leftarrow U_\ibold^{n+1}$
  and $B_{\ibold+\half\edbold}^n\leftarrow
  B_{\ibold+\half\edbold}^{n+1}$ and synchronize cell-centered and
  face-centered magnetic field values.
\item update solution time $t\leftarrow t+\Delta t$ and compute new
  timestep according to the Courant-Friedrichs-Lewy (CFL) condition
  for stability,
 \begin{equation}
   \Delta t\leftarrow C_{\rm CFL}\frac{\Delta
    x}{\mbox{Max}(u^{n+1}_{\ibold,d}+c^{n+1}_{f,\ibold})},
  \end{equation}
  where the
  CFL number is, $C_{\rm CFL}<1$, $u_d$ is the $d-$component of the
  velocity field, $c_f$ is the fast magnetosonic wave speed and the
  max value is computed over all directions $d$ and over all cells in
  the computational domain.
\end{enumerate}
In the following subsections we provide additional details about the
algorithm.

\subsubsection{Primitive Variables} \label{primitive:se}

The first part of the algorithm consists in reconstructing the
numerical solution within the spatial domain of the mesh.  This
requires estimating gradients and eventually extrapolating 
state variables in time and space from cell centers to cell faces with
the use of characteristic tracing. Such operations are usually done in
{\it primitive} space, where the variables are defined as
\begin{equation} \label{primitives:eq}
W=\left(\rho,u_0,\dots,u_{\Dim-1},p_g,B_0,\dots,B_{\Dim-1}\right)^T,
\end{equation}
as it simplifies the characteristic analysis of the system.  The
evolution of the system in terms of primitive variables is given by a
system of equations in non-conservative form,
\begin{gather} \label{prim_sys:eq}
\frac{\partial W}{\partial t}
+ \sum^{\Dim-1}_{d=0} A_d(W) ~ \frac{\partial W_d}{\partial x_d}
= S^W,
\end{gather}
where $S^W = \nabla_U W \cdot S$ and $A_d = \nabla_U W \cdot \nabla_U
F_d \cdot \nabla_W U \notag$, and $\nabla_U W$ and $\nabla_U W$ are
the Jacobian of the transformation from primitive to conserved
variables and vice-versa.  Given their importance for the construction
of a sound numerical scheme, the properties of the operators $A_d$
have been studied in great details in the
literature~\citep[e.g.][]{BrioWu1988JCoPh..75..400B,RoeBalsara1996}.
Here it suffices to make the following observations.  In one dimension,
$d=0$, the parallel component of the magnetic field is constant,
$B_0=const.$, and $A_0$ effectively becomes a $7\times 7$ matrix. In
general the operator $A_0$ is characterized by 7 eigenvalues, and
associated left and right eigenvectors, with values $u$, $u\pm c_s$,
$u\pm c_A$ and $u\pm c_f$ obeying the hierarchy: $c_s\le c_A \le
c_f$.  The first eigenvalue listed above corresponds to the usual
entropy wave, and the other six to the three pairs of MHD waves (slow
magnetosonic, Alfv\'en and fast magnetosonic) propagating downstream (+)
or upstream (-) in the flow, respectively.  Because up to 5 of the
eigenvalues may actually coincide, the system is not strictly
hyperbolic, so care must be taken to avoid singularities with
expressions involving the
eigenvectors~\citep{BrioWu1988JCoPh..75..400B,RoeBalsara1996}.
Finally, in more then one dimensions, because $B_d$ is not affected by
gradients in the $d$ direction, it has been customary to use the
one-dimensional analysis when formulating the predictor step for
higher-order Godunov-like MHD algorithms.  However, that leads to
neglect of terms $\propto\partial B_d/\partial x_d$, which are not
necessarily null in multidimensions. In fact, it is important to
include these terms to avoid degrading the solution
accuracy, as recently pointed out
by~\cite{GardinerStone2005JCoPh.205..509G}.

\subsubsection{Slopes} \label{slopes:se}

After the conversion from conservative to primitive
variables, $U^n_\ibold\leftarrow W^n_\ibold(U^n_\ibold)$ and
the synchronization of cell-centered and face-centered magnetic fields
according to Eq. (\ref{bcc:eq}), we proceed to the calculation of
the slopes along each direction $d$ as follows.  First, central and
side slopes are estimated, respectively, as
\begin{align} \label{cslope:eq}
\delta_{d,0} W_\ibold = & \half\left(W^n_{\ibold + \ebold^d} - W^n_{\ibold - \ebold^d}\right), \\
\delta_{d,-} W_\ibold= & W^n_\ibold - W^n_{\ibold - \ebold^d},\\
\delta_{d,+} W_\ibold= & W^n_{\ibold + \ebold^d} - W^n_\ibold, \label{rslope:eq}
\end{align}
and then {\it limiting} is applied component-wise either in primitive
or in characteristic space. We use van Leer's limiter defined as
\begin{align*}
\delta^{vL}(\delta_0,\delta_-,\delta_+) =
  \begin{cases}
\mbox{sgn}(\delta_0)\; \min(|\delta_0|,2|\delta_-|,2|\delta_+|) & \mbox{if $\delta_-\delta_+ > 0$} \\
  0 &    \text{otherwise}.
  \end{cases}
\end{align*}
In the case of primitive limiting the limiter is applied directly to
each component $k$ of the primitive slopes
(\ref{cslope:eq})-(\ref{rslope:eq}), i.e.
\begin{equation}
\delta_d W_\ibold^k=\delta^{vL}(\delta_{d,0}W_\ibold^k,\delta_{d,-}W_\ibold^k,\delta_{d,+} W_\ibold^k).
\end{equation}
In the case of characteristic limiting the limiter is applied to
the components of the primitive slopes in characteristic space, namely
\begin{eqnarray}
\delta_d W_\ibold &=& \sum_k \alpha^kr^k ,\\
\alpha^k&=&\delta^{vL}(\alpha^k_0,\alpha^k_-,\alpha^k_+), \\
\alpha^k_\# &=& l^k \cdot \delta_{d,\#} W_\ibold,\quad\#=0,+,-,
\end{eqnarray}
where, $l^k= l^k(W^n_\ibold)$ and $r^k = r^k(W^n_\ibold)$ are the left
and right eigenvectors of the operator $A_{\ibold,d}$.

\subsubsection{Normal Predictor} \label{nps:se}

Next we extrapolate the primitive variables from cell centers to face
centers along each coordinate direction, $d$, by taking into account
the time-averaged effect due to the slopes just computed.  This is
done most conveniently by 
using the 1-D versions of the MHD equations in primitive form.
There is no evolution in the $d-$component of the magnetic field due
to derivatives along the $d-$direction, so the normal components of
the magnetic field on cell faces are straightforwardly provided by the
face-centered component of the magnetic field, without need for even
geometrical extrapolation.  On the other hand, as pointed out 
by~\cite{GardinerStone2005JCoPh.205..509G}
in multidimensional MHD the reconstruction
step must include terms proportional to $\partial B_d/\partial x_d$
terms (no summation over repeated indices is implied here). These terms
arise from the requirement to balance the divergence terms in more
than one dimension and their neglect can cause serious degradation of the
numerical solution.
The reconstruction of the primitive variables onto cell faces can be
done with various degrees of accuracy. Typically, one uses a
piecewise-linear (PLM) or piecewise-parabolic (PPM) reconstruction 
scheme~\citep{colellawoodward84}. 
Although we mostly use a PPM algorithm for the tests presented
here, for simplicity we illustrate the case of PLM reconstruction.
So in this case the extrapolation of the primitive variables in space and
time from cell centers to faces along the direction $d$ takes the form
\begin{alignat}{1} 
\begin{pmatrix}
\hat W^{n}_{\ibold, \pm, d} \\ B^{n}_{d,\ibold,\pm,d}
\end{pmatrix}
= \; 
\begin{pmatrix}
\hat W^n_\ibold  \\ B^{n}_{d,\ibold+\edbold}
\end{pmatrix} 
 +\half \left[
\begin{pmatrix}
       \pm I & 0 \\ 0 & 0
\end{pmatrix}
-\frac{\Delta t}{\Delta x}
\begin{pmatrix}
        \hat A_{\ibold,d} \\ 0
\end{pmatrix}
P_\pm (\delta_d \hat W^n_\ibold )
\right] 
+\frac{\Delta t}{2}
\begin{pmatrix}
        S_{d,\ibold}^{n,\rm MHD} \\ 0 
\end{pmatrix}.
\label{normalpred:eq} 
\end{alignat}
where we have explicitly separated out the reconstruction of the 
normal component of the magnetic field (the {\it hat} symbol in the notation
indicates that the components corresponding to the normal magnetic
field are omitted). In addition, we have used the projector operator
defined as
\begin{equation}
P_\pm(W) = \sum_{\pm \lambda_k > 0} (l_k \cdot W) r_k
\end{equation}
where $\lambda_k$ are eigenvalues of $A_{\ibold,d}$. This projector operator
filters out the components of the gradients that propagate away from
the cell interface. However, when a Riemann solver of the HLL family
is employed, in order to obtain second-order accuracy 
the filter is switched off and both in the PPM and PLM cases, 
and the summation is carried over all waves, irrespective of their sign
(this is further discussed in Sec.\ref{rs:sec}).
Finally, $S_{d,\rm MHD}$ represents the MHD source
term required in multidimensional MHD which we
implement in the form~\citep{Crockett2005JCoPh.203..422C}
\begin{equation}
S_{d,\ibold}^{n,\rm MHD}=
\begin{pmatrix}
0\\
\frac{B_{0}}{\rho} \\
\vdots \\
\frac{B_{\Dim-1}}{\rho} \\
{\bf B\cdot u}\\
u_{d_1} \\
u_{d_2}
\end{pmatrix}_\ibold^n \left(\frac{\partial B_d}{\partial x_d}\right)_\ibold^n
\end{equation}
where, as usual, $d\ne d_1\ne d_2$ and, in this particular case,
$0\le d_1<d_2<\Dim$.
The normal predictor step is completed by the final corrections for a
non-stiff source term according to
\begin{equation}
W^{n}_{\ibold, \pm, d}= W^{n}_{\ibold, \pm, d} +\frac{\Delta t}{2} S^{n}_\ibold.
\end{equation}

\subsubsection{CT Extended Corner Transport Upwind} \label{ctu:se}

After the primitive variables have been extrapolated to cell faces, we
add corrections due to gradients parallel to the cell faces. We find it 
convenient at this point
to convert back to a conservative representation, thus we operate the
transformation: $W^{n}_{\ibold,\pm,d}\leftarrow U^{n}_{\ibold,\pm,d}$.
Following the CTU scheme the corrections are 
expressed in terms of transverse flux gradients, with fluxes obtained 
from a Riemann solver, ${\cal R}$. In accord with the
CT scheme, however, for the
update of magnetic field variables we  use 
gradients of edge-centered electric fields 
suitably interpolated in space and time from their 
face and cell centered values. Both the interpolation procedure,
${\cal I}_E$, and the Riemann solver, ${\cal R}$, will be specified
at the end of this section. It suffices here to say that 
the time centering and interpolation accuracy
of the interpolated edge-centered electric fields is consistent 
with that of the MHD fluxes returned by the Riemann solver.

The steps involved in the modified CTU update can then be summarized as follows:

\begin{enumerate}
\item 
Use a Riemann solver, ${\cal R}(U^{Left},U^{Right})$, to 
obtain a first estimate of the fluxes across cell faces along each 
direction, $d$
\begin{equation} \label{riemann1:eq}
F^{\text{1D}}_{d,\ibold + \half \edbold} = 
 {\cal R} (U^n_{\ibold, +, d}, U^n_{\ibold + \edbold, -, d},d).
\end{equation}

\item Use the newly obtained fluxes with the primitive solution at time $t^n$
to interpolate the electric fields from cell-faces and cell-centers to cell-edges
\begin{equation} \label{einterp1:eq}
E^{\text{1D}}_{d,\ibold+\half\eboldone+ \half\eboldtwo} =
{\cal I}_E(F^{\text{1D}}_{d_1,* + \half \eboldone},
F^{\text{1D}}_{d_2,*+\half\eboldtwo},W^{n}_{*}),
\end{equation}
where the $*$ symbol indicates that the interpolation requires
values of the arguments at various cell centers and faces.

\item 
As part of the CTU prescription to obtain $(1,1,1)$ diagonal 
coupling, apply corrections to density, momentum and energy components 
of $U_{\ibold, \pm, d_1}$, due to one set of transverse flux derivatives
along $d_2$. For each face $d_1$, there will be $\Dim-1$ such corrected
states, one for each direction perpendicular to $d_1$. 
The corrected states read
\begin{equation} \label{uupdate1:eq}
U^n_{\ibold, \pm, d_1, d_2} = U^n_{\ibold, \pm, d_1}
    - \frac{\Delta t}{3\Delta x} 
      (F^{\text{1D}}_{d_2,\ibold + \half \eboldtwo}
     - F^{\text{1D}}_{d_2,\ibold - \half \eboldtwo}).
\end{equation}

\item Likewise, use the CT scheme to correct magnetic field components 
affected by the same set of transverse flux derivatives.
There are $\Dim-1$ magnetic field components of $U^n_{\ibold, \pm, d_1}$ 
on the face $d_1$ which are affected by the transverse
flux along $d_2$. First, the component along $d_1$, which we 
indicate with $U_{B_{d_1}}$, is corrected as
\begin{equation}
\label{b1update1:eq}
U_{B_{d_1},\ibold \pm,d_1,d_2}^{n} = 
U_{B_{d_1},\ibold \pm,d_1}^{n}
-\varepsilon_{d_1d_3d_2} \frac{\Delta t}{3\Delta x}
\left(
E^{\text{1D}}_{d_3,\ibold \pm\half\eboldone+\half\eboldtwo} - 
E^{\text{1D}}_{d_3,\ibold \pm\half \eboldone-\half\eboldtwo}\right).
\end{equation}
Second, we correct the magnetic field component, $U_{B_{d_3}}$, parallel to the cell
face and directed along the remaining direction $d_3\ne d_2\ne d_1$.
As in~\cite{GardinerStone2008JCoPh.227.4123G},
for this component we average the contributions from the faces at
$\ibold +\half\ebold^{d_3}$ and $\ibold -\half\ebold^{d_3}$,
obtaining
\begin{eqnarray}
\notag
U_{B_{d_3},\ibold \pm,d_1,d_2}^{n} =  
U_{B_{d_3},\ibold \pm,d_1}^{n}
-\varepsilon_{d_1d_3d_2} \frac{\Delta t}{6\Delta x}
\left[\left(
E^{\text{1D}}_{d_1,\ibold +\half\ebold^{d_3}+\half\eboldtwo} - 
E^{\text{1D}}_{d_1,\ibold +\half\ebold^{d_3}-\half\eboldtwo}\right)\right.
\\
+\left. \left(
E^{\text{1D}}_{d_1,\ibold -\half\ebold^{d_3}+\half\eboldtwo} - 
E^{\text{1D}}_{d_1,\ibold -\half\ebold^{d_3}-\half\eboldtwo}\right)
\right].
\label{b3update1:eq}
\end{eqnarray}

\item
 Use a Riemann  solver  to obtain  fluxes  for each  pair of  states
 corrected for transverse fluxes. This provides $\Dim-1$ fluxes per
cell face.
\begin{equation} \label{riemann2:eq}
F_{d_1,\ibold + \half \ebold^{d_1}, d_2} =
{\cal R} (U_{\ibold, +, d_1, d_2}, U_{\ibold + \ebold^{d_1}, -, d_1, d_2}, d_1)
\end{equation}
\begin{equation*}
d_1 \neq d_2, ~ 0 \le d_1, d_2 < \Dim
\end{equation*}
\item Obtain new interpolated values of the electric field 
from the averages of the above computed fluxes
\begin{equation} \label{einterp2:eq}
E_{d,\ibold+\half\eboldone+ \half\eboldtwo} =
{\cal I}_E(\tilde F_{d_1,* + \half \eboldone},
\tilde F_{d_2,*+\half\eboldtwo},W^{n}_{*})
\end{equation}
where
\begin{eqnarray} \label{flxtld:eq}
\tilde F_{d_1,\ibold + \half \eboldone}&= &\half
(F_{d_1,\ibold + \half \eboldone,d_2}+F_{d_1,\ibold + \half \eboldone,d_3}) \\
\tilde F_{d_2,\ibold + \half \eboldtwo}& = &\half
(F_{d_2,\ibold + \half \eboldtwo,d_1}+F_{d_2,\ibold + \half \eboldtwo,d_3}) .
\end{eqnarray}

\item Compute final corrections to the density, momentum and energy 
components of $U_{\ibold, \pm, d}$ due to transverse fluxes using
the above Riemann solutions, according to
\begin{alignat}{2} \notag
 U^{n+\half}_{\ibold, \pm, d} = \; & U_{\ibold, \pm, d}
 \; & & - \; \frac{\Delta t}{2\Delta x}
                         (F_{d_1,\ibold + \half \ebold^{d_1}, d_2}
                        - F_{d_1,\ibold - \half \ebold^{d_1}, d_2}) \\
 & \; & & - \; \frac{\Delta t}{2\Delta x}
                        (F_{d_2,\ibold + \half \ebold^{d_2}, d_1}
                       - F_{d_2,\ibold - \half \ebold^{d_2}, d_1}) \\
 & & & d \neq d_1 \neq d_2, ~ 0 \le d, d_1, d_2 < \Dim \notag
\end{alignat}

\item Likewise, use the CT scheme to apply final corrections
to magnetic field components. In analogy to Eq. (\ref{b1update1:eq})
in step 4 we write
\begin{eqnarray}\notag
U_{B_d,\ibold,\pm,d}^{n+\half} = U_{B_d,\ibold,\pm,d}^{n}
-\varepsilon_{dd_1d_2} \frac{\Delta t}{2\Delta x}
\left[\left(
E_{d_1,\ibold\pm\half\edbold+\half\eboldtwo} - 
E_{d_1,\ibold \pm\half\edbold-\half\eboldtwo}\right)\right. \\
-\left.\left(
E_{d_2,\ibold\pm\half\edbold+\half\eboldone} - 
E_{d_2,\ibold \pm\half\edbold-\half\eboldone}\right)\right]
\label{b1update2:eq}
\end{eqnarray}
Finally, in analogy to Eq. (\ref{b3update1:eq})
the magnetic field components parallel to the cell
face are updated according to the CT scheme as
\begin{eqnarray}
\notag
U_{B_{d_1},\ibold \pm,d}^{n+\half} =  
U_{B_{d_1},\ibold \pm,d}^{n}
-\varepsilon_{dd_1d_2} \frac{\Delta t}{2\Delta x}
\left[\left(
E_{d,\ibold +\half\ebold^{d_1}+\half\eboldtwo} - 
E_{d,\ibold +\half\ebold^{d_1}-\half\eboldtwo}\right)\right. \\
\notag
+\left(
E_{d,\ibold -\half\ebold^{d_1}+\half\eboldtwo} - 
E_{d,\ibold -\half\ebold^{d_1}-\half\eboldtwo}\right) 
\\
\notag
-\left(
E_{d_2,\ibold +\half\ebold^{d_1}+\half\edbold} - 
E_{d_2,\ibold +\half\ebold^{d_1}-\half\edbold}\right) \\
-\left.
 \left(
E_{d_2,\ibold -\half\ebold^{d_1}+\half\edbold} - 
E_{d_2,\ibold -\half\ebold^{d_1}-\half\edbold}\right)
\right].
\label{b3update2:eq}
\end{eqnarray}

\item Compute final second-order estimate of time averaged fluxes
\begin{equation} \label{riemann3:eq}
F^{n+\half}_{d,\ibold + \half \ebold^d} =
     {\cal R} (U^{n+\half}_{\ibold, +, d},
             U^{n+\half}_{\ibold + \ebold^d, -, d}, d).
\end{equation}
\item Compute final estimate of the electric field
using the above fluxes and an updated 
value for the cell-centered conservative variables
using the averaged fluxes in Eq. (\ref{flxtld:eq}),
namely
\begin{equation} \label{einterp3:eq}
E^{n+\half}_{d,\ibold+\half\eboldone+ \half\eboldtwo} =
{\cal I}_E(F^{n+\half}_{d_1,* + \half \eboldone},
F^{n+\half}_{d_2,*+\half\eboldtwo},\tilde W^{n}_{*}),
\end{equation}
where
\begin{equation*}
\tilde W_{\ibold}^n= W_{\ibold}^n - 
\half\frac{\Delta t}{\Delta x}\;
\nabla_UW\cdot\sum\limits^{\Dim - 1}_{d=0}\left(
 \tilde F_{d,\ibold + \half \ebold^d} -
 \tilde F_{d,\ibold - \half \ebold^d} \right).
\end{equation*}
and the operator $\nabla_UW$ symbolizes the transformation
from conservative to primitive variables.
\item Finally update cell-centered conservative variables using
  Eq. (\ref{cons_fv:eq}) with the fluxes in (\ref{riemann3:eq}) and
  the face-centered magnetic field variables using
  Eq. (\ref{faraday_fa:eq}) with the electric field in
  (\ref{einterp3:eq}) and synchronize the magnetic variables using
  Eq. (\ref{bcc:eq}).
\end{enumerate}

\subsubsection{Riemann Solver} \label{rs:sec}
For the purpose of this paper, 
we have implemented the HLLD solver recently developed
by~\cite{MiyoshiKusano2005JCoPh.208..315M}.  This is an extended
version of the original HLL solver for the MHD, which includes the
entropy, Alfv\'en and fast magnetosonic waves.  The solver appears
quite accurate and robust and relatively inexpensive.  As already
pointed out in order for the fluxes returned by HLL-type solvers to be
second-order accurate, the projector $P$ is modified in such a way
that the summation is carried over all waves, irrespective of their
sign. This is because the fluxes returned by these solvers are
built on the spatially reconstructed solutions 
on the left and right hand sides of the cell interfaces.
The solver is extensively documented in the original paper and
its description will not be repeated here.

\subsubsection{Interpolation Scheme for the Edge-Centered Electric Field}
\label{ecef:sec}

In this section we describe the scheme used to interpolate the face-centered
electric fields returned by the Riemann solver onto cell 
edges. In Sec. \ref{ctu:se} this was indicated with the notation,
${\cal I}_E$.  It is important for the stability of the overall
algorithm to choose this interpolation scheme in
such a way that there is consistency between the time-update of the
face-centered magnetic field, the cell-centered variables and the
synchronization step of the magnetic variables, Eq.~(\ref{bcc:eq}).
For this purpose we have adopted the upwind scheme described
in~\cite{GardinerStone2005JCoPh.205..509G,GardinerStone2008JCoPh.227.4123G}.
This will be described next, for the case of three dimensions.  In
this case each cell edge is shared by four adjacent faces from which
the electric field can be interpolated. Hence, these four interpolated
values will be arithmetically averaged.  The scheme for ${\cal
  I}_E()$ uses three arguments,
$$E_{d,\ibold+\half\eboldone+\half\eboldtwo} ={\cal I}_E(F_{d_1,*+\half\eboldone},F_{d_2,*+\half\eboldtwo},W_*),$$
where the $*$ symbol indicates that the interpolation requires
values of the arguments at different cells and faces.
The face-centered fluxes define the
face-centered electric fields according to Eq.~(\ref{fcef:eq}), and the
cell-centered primitive state variable, $W_\ibold$, is used to define a
cell-centered electric field according to the usual formula
$$E_{d,\ibold}=-\varepsilon_{dd_1d_2}(u_{d_1,\ibold}B_{d_2,\ibold}-u_{d_2,\ibold}B_{d_1,\ibold}),$$
with the velocity and magnetic field variables given by the components
of $W_\ibold$. The cell-centered electric field is used together
with face-centered electric fields~(\ref{fcef:eq}) to define the following
transverse quasi-cell-centered gradients
\begin{equation}\label{qccg:eq}
\left(\frac{\delta E_d}{\delta x_{d_1}}\right)_{\ibold+\fourth\eboldone} =
2\frac{E_{d,\ibold+\half\eboldone}-E_{d,\ibold}}{\Delta x}.
\end{equation}
In turn, these gradients are used to define quasi-face-centered gradients
of the electric field by the upwind 
scheme~\citep{GardinerStone2005JCoPh.205..509G}
\begin{align*}
\left(\frac{\delta E_d}{\delta x_{d_2}}\right)_{\ibold+\half\eboldone+\fourth\eboldtwo} =
  \begin{cases}
\left(\frac{\delta E_d}{\delta x_{d_2}}\right)_{\ibold+\fourth\eboldtwo} & \mbox{if $u_{\ibold+\half\eboldtwo}> 0$} \\
\left(\frac{\delta E_d}{\delta x_{d_2}}\right)_{\ibold+\frac{3}{4}\eboldtwo} & \mbox{if $u_{\ibold+\half\eboldtwo}< 0$} \\
\half\left[\left(\frac{\delta E_d}{\delta x_{d_2}}\right)_{\ibold+\fourth\eboldtwo}+\left(\frac{\delta E_d}{\delta x_{d_2}}\right)_{\ibold+\frac{3}{4}\eboldtwo}\right] &    \text{otherwise}.
  \end{cases}
\end{align*}
With the above definitions, the interpolated edge-centered 
electric field is defined as
\begin{eqnarray}  \notag
E_{d,\ibold+\half\eboldone+\half\eboldtwo} = 
&\frac{1}{4}&
\left(
E_{d,\ibold+\half\eboldone} 
+E_{d,\ibold+\eboldtwo+\half\eboldone}
+E_{d,\ibold+\half\eboldtwo}
+E_{d,\ibold+\eboldone+\half\eboldtwo}\right) \\\label{eedge:eq}
+&\frac{\Delta x}{8}&
\left[
\left(\frac{\delta E_d}{\delta x_{d_2}}\right)_{\ibold+\half\eboldone+\fourth\eboldtwo}
-
\left(\frac{\delta E_d}{\delta x_{d_2}}\right)_{\ibold+\half\eboldone+\frac{3}{4}\eboldtwo}\right]
\\ \notag
+&\frac{\Delta x}{8}&
\left[
\left(\frac{\delta E_d}{\delta x_{d_1}}\right)_{\ibold+\half\eboldtwo+\fourth\eboldone} 
-
\left(\frac{\delta E_d}{\delta x_{d_1}}\right)_{\ibold+\half\eboldtwo+\frac{3}{4}\eboldone} 
\right].
\end{eqnarray}

\section{Tests} \label{tests:se}

\subsection{Convergence Rates in Smooth Flows} \label{convrate:se}
In this section we test the correctness of our implementation by
measuring the convergence rate of the numerical solution returned by
the code.  The test is based on the propagation of Alfv\'en, fast and
slow MHD waves. The waves have small amplitude, $\delta=10^{-5}$, so
that we expect to observe the nominal second order convergence rate
predicted by numerical analysis.  

In the following tests the initial conditions are provided
for the primitive variables by defining the unperturbed state, $W$,
and the superposed perturbation, $\delta W$, corresponding to the
wave.
The size of the computational box is, $L=1$, the geometry is
one- or two-dimensional and the boundary conditions are periodic.
The adiabatic index is $\gamma=5/3$. While
we have experimented with different choices of orientation of the
wave-vector with respect to the grid, namely ${\bf k} / 2\pi = (1,0),~
(0,1), ~(1/\sqrt{2},1/\sqrt{2}), ~(2/\sqrt{5},1/\sqrt{5})$, 
we find the same convergence rates in all cases. Thus, we simply report 
the results for the few representative tests
listed in Table~\ref{runset:tab}.
\begin{table}  
\caption{Run Set\label{runset:tab}}
\begin{tabular*}{\textwidth}{@{\extracolsep{\fill}}lccc}
\hline \hline 
run & ${\delta}$ & ${\bf k}/2\pi$ &  Type  \cr
\hline
A& $10^{-5}$ & $(1,0)$  &  {\rm Alfv\'en}   \cr  
B& $10^{-5}$ & $(2/\sqrt{5},1/\sqrt{5})$  &  {\rm Alfv\'en}   \cr  
C& $10^{-5}$ & $(1,0)$  &  {\rm Fast}   \cr  
D& $10^{-5}$ & $(2/\sqrt{5},1/\sqrt{5})$  &  {\rm Fast}   \cr  
E& $10^{-5}$ & $(1,0)$  &  {\rm Slow}   \cr  
F& $10^{-5}$ & $(2/\sqrt{5},1/\sqrt{5})$  &  {\rm Slow}   \cr  
\hline
\end{tabular*}
\end{table}

In order to measure the rate at which the numerical solution
converges, for each problem we carry out a set of 5 simulation runs
employing $N_{cell}=16,32,64,128,256$ for a total range of 16.
For each run the time-step size is fixed, scales inversely with
$N_{cell}$, and corresponds roughly to a CFL number 0.8. The
convergence rate is measured using Richardson extrapolation.
Given the numerical solution $q_{r}$ at resolution $r$ 
we first estimate the error at a given point $(i,j)$, as
\begin{equation} \label{numerr:eq}
\varepsilon_{r;i,j} = q_{r}(i,j) - \bar q_{r+1}(i,j),
\end{equation}
where $\bar q_{r+1}$ is the solution at the next finer resolution,
spatially averaged onto the coarser grid.
We then take the n-norm of the error
\begin{equation} \label{lnorm_n:eq}
L_n = \| \varepsilon_{r} \|_n =  \left( \sum |\varepsilon_{r;i,j}|^n  v_{i,j}\right)^{1/n},
\end{equation}
where, $v_{i,j}=\Delta x^2$ is the cell volume, 
and estimate the convergence rate as
\begin{equation}
R_n = \frac{ \ln[L_n(\varepsilon_r)/L_n(\varepsilon_s)] }{ \ln (\Delta x_r / \Delta x_s) }.
\end{equation}
For each case listed in Table~\ref{runset:tab}, we produce a
corresponding Table~\ref{alfven:tab}-\ref{slow:tab} reporting the
$L_1,~L_2$ and $L_\infty$ norms of the error and the corresponding
convergence rates, $R_1,~R_2$ and $R_\infty$, as defined above.

\subsubsection{Alfv\'en Waves}

The test based on the propagation of 
the Alfv\'en wave is characterized by the following
unperturbed state and superposed 
perturbation~\citep{Crockett2005JCoPh.203..422C}
\begin{equation}
W=
\begin{pmatrix}
\rho_0 \\
u_x \\
u_y \\
u_z \\
P_0 \\
B_0/\sqrt{2} \\
B_0/\sqrt{2} \\
0 \\
\end{pmatrix}, \quad \quad
\delta W =
\begin{pmatrix}
0 \\
0 \\
0 \\
-c_A\\
0\\
0\\
0\\
B_0\\
\end{pmatrix} \delta \sin({\bf k \cdot x})
\label{alfven:ics}
\end{equation}
where $\rho_0=P_0=B_0=1$ and $u_x=u_y=u_z=0$,
$c_A=B_0/\sqrt{\rho_0}=1$, is the Alfv\'en speed, and ${\bf k}$
and ${\bf x}$ are the wavevector and position vector respectively.  The
convergence rates for the perturbed quantities are summarized in Table
\ref{alfven:tab}. In both tests, with different ${\bf k}$,
both the $z-$components of the
velocity and magnetic field converge with second-order accuracy. The
error on the unperturbed variables (not reported) is much smaller and
converges at the same rate as the perturbed variables until
dominated by the machine round-off error. As already pointed out, 
tests with different orientation of $\bf k$ show the same
convergence rates.
%
%
%
\begin{table}  
\begin{small}
\caption{Convergence Rates for Alfv\'en Waves.\label{alfven:tab}}
\begin{tabular*}{\textwidth}{@{\extracolsep{\fill}}lcccccccccccc}
\hline
\hline \cr
N$_{\rm cells}$ & $L_1$ & $R_1$ & $L_2$ & $R_2$ & $L_\infty$ & $R_\infty$ & $L_1$ & $R_1$ & $L_2$ & $R_2$ & $L_\infty$ & $R_\infty$ \cr \cline{2-13}
\cline{2-7} \cline{8-13} \cr
&\multicolumn{12}{c}{Case A: $\delta= 10^{-5},~{\bf k}=2\pi(1,0)$} \cr
& \multicolumn{6}{l}{\bf $z-$velocity} & \multicolumn{6}{l}{\bf $z-$magnetic} \cr
16 &       7.5E-09 &    --  &    1.7E-08 &    --  &    4.7E-08 &    --  &    7.5E-09 &    --  &    1.7E-08 &    --  &    4.7E-08 &    -- \cr 
32 &       1.9E-09 &    2.0 &    4.2E-09 &    2.0 &    1.2E-08 &    2.0 &    1.9E-09 &    2.0 &    4.2E-09 &    2.0 &    1.2E-08 &    2.0 \cr 
64 &       4.8E-10 &    2.0 &    1.1E-09 &    2.0 &    3.0E-09 &    2.0 &    4.8E-10 &    2.0 &    1.1E-09 &    2.0 &    3.0E-09 &    2.0 \cr 
128&       1.2E-10 &    2.0 &    2.7E-10 &    2.0 &    7.5E-10 &    2.0 &    1.2E-10 &    2.0 &    2.7E-10 &    2.0 &    7.5E-10 &    2.0  \cr
\cline{2-13}

\cr
&\multicolumn{12}{c}{Case B: $\delta= 10^{-5},~{\bf k}=2\pi(2,1)/\sqrt{5}$} \cr
& \multicolumn{6}{l}{\bf $z-$velocity} & \multicolumn{6}{l}{\bf $z-$magnetic} \cr
   16 &  7.6E-08 &    --  &    1.2E-07 &    --  &    2.4E-07 &    --  &    7.8E-08 &    --  &    1.2E-07 &    --  &    2.3E-07 &    --  \cr 
   32 &  2.0E-08 &    1.9 &    3.1E-08 &    2.0 &    6.2E-08 &    1.9 &    2.0E-08 &    2.0 &    3.1E-08 &    2.0 &    6.2E-08 &    1.9 \cr 
   64 &  4.9E-09 &    2.0 &    7.8E-09 &    2.0 &    1.5E-08 &    2.0 &    5.0E-09 &    2.0 &    7.8E-09 &    2.0 &    1.6E-08 &    2.0 \cr 
  128 &  1.2E-09 &    2.0 &    1.9E-09 &    2.0 &    4.0E-09 &    2.0 &    1.2E-09 &    2.0 &    2.0E-09 &    2.0 &    4.0E-09 &    2.0  \cr 
\hline
\hline
\end{tabular*}
\end{small}
\end{table}

\subsubsection{Fast and Slow Magnetosonic Waves}
For fast and slow magnetosonic waves the unperturbed 
state and the superposed perturbations 
read~\citep{Crockett2005JCoPh.203..422C}
\begin{equation}
W=
\begin{pmatrix}
\rho_0 \\
u_x \\
u_y \\
u_z \\
P_0 \\
B_0 \hat b_x\\
B_0 \hat b_y\\
0 \\
\end{pmatrix}, \quad \quad
\delta W =
\begin{pmatrix}
\rho_0 \\
(\sqrt{2}c^2_w\hat b_y -c_g^2 \hat k_y)/c_w\\
-(\sqrt{2}c^2_w\hat b_x-c_g^2\hat k_x)/c_w\\
0 \\
\rho_0c_g^2\\
-\sqrt{2}B_0(c_w^2-c_g^2)\hat k_y/c_A^2\\
\sqrt{2}B_0(c_w^2-c_g^2)\hat k_x/c_A^2\\
0\\
\end{pmatrix} \delta \sin({\bf k \cdot x})
\label{fastslow:ics}
\end{equation}
where, $\hat b$ is the unit vector along the magnetic field vector and
is at $\pi/4$ radians with respect to the unit vector 
$\hat k\equiv{\bf k}/k$, $c_g=\sqrt{\gamma P_0/\rho_0}$ is the 
gas sound speed, $c_w$ is the speed of fast or slow MHD waves and 
all other symbols take the
meaning and values as in the previous section.

The convergence rates are reported in
Table~\ref{fast:tab} and~\ref{slow:tab} for fast and slow waves,
respectively.
As with Alfv\'en waves, the errors in the perturbed variables 
converge with second order accuracy, while the error in the unperturbed
variables are much smaller and converge at the same rate until
they are affected by machine precision.

\begin{table}  
\begin{small}
\caption{Convergence Rates for Fast MHD Waves.\label{fast:tab}}
\begin{tabular*}{\textwidth}{@{\extracolsep{\fill}}lcccccccccccc}
\hline
\hline \cr
N$_{\rm cells}$ & $L_1$ & $R_1$ & $L_2$ & $R_2$ & $L_\infty$ & $R_\infty$ & $L_1$ & $R_1$ & $L_2$ & $R_2$ & $L_\infty$ & $R_\infty$ \cr \cline{2-13} \cr
&\multicolumn{12}{c}{Case C: $\delta= 10^{-5},~{\bf k}=2\pi(1,0)$} \cr
& \multicolumn{6}{l}{\bf density-gas} & \multicolumn{6}{l}{\bf $x-$vel-gas} \cr
   16 &    7.5E-09 &    --  &    1.7E-08 &    --  &    4.6E-08 &    --  &    1.1E-08 &    --  &    2.4E-08 &    --  &    6.6E-08 &    --  \cr 
   32 &    1.9E-09 &    2.0 &    4.2E-09 &    2.0 &    1.2E-08 &    2.0 &    2.7E-09 &    2.0 &    6.0E-09 &    2.0 &    1.7E-08 &    2.0 \cr 
   64 &    4.8E-10 &    2.0 &    1.1E-09 &    2.0 &    3.0E-09 &    2.0 &    6.8E-10 &    2.0 &    1.5E-09 &    2.0 &    4.3E-09 &    2.0 \cr 
  128 &    1.2E-10 &    2.0 &    2.7E-10 &    2.0 &    7.5E-10 &    2.0 &    1.7E-10 &    2.0 &    3.8E-10 &    2.0 &    1.1E-09 &    2.0 \cr 
\cline{2-7} \cline{8-13} \cr
& \multicolumn{6}{l}{\bf $y-$vel-gas} & \multicolumn{6}{l}{\bf pressure} \cr
   16 &    3.4E-09 &    --  &    7.6E-09 &    --  &    2.1E-08 &    --  &  1.0E-08 &    --  &    2.3E-08 &    --  &    6.5E-08 &    --   \cr 
   32 &    8.7E-10 &    2.0 &    1.9E-09 &    2.0 &    5.5E-09 &    2.0 &  2.7E-09 &    2.0 &    5.9E-09 &    2.0 &    1.7E-08 &    2.0   \cr 
   64 &    2.2E-10 &    2.0 &    4.9E-10 &    2.0 &    1.4E-09 &    2.0 &  6.7E-10 &    2.0 &    1.5E-09 &    2.0 &    4.2E-09 &    2.0   \cr 
  128 &    5.5E-11 &    2.0 &    1.2E-10 &    2.0 &    3.5E-10 &    2.0 &  1.7E-10 &    2.0 &    3.7E-10 &    2.0 &    1.1E-09 &    2.0   \cr 
\cline{2-7} \cline{8-13} \cr
& \multicolumn{6}{l}{\bf $y-$magnetic} & \multicolumn{6}{l}{} \cr
   16 &    7.0E-09 &    --  &    1.5E-08 &    --  &    4.3E-08 &    --  &  & & & & &   \cr 
   32 &    1.8E-09 &    2.0 &    3.9E-09 &    2.0 &    1.1E-08 &    2.0 &  & & & & &     \cr 
   64 &    4.4E-10 &    2.0 &    9.9E-10 &    2.0 &    2.8E-09 &    2.0 &  & & & & &     \cr 
  128 &    1.1E-10 &    2.0 &    2.5E-10 &    2.0 &    7.0E-10 &    2.0 &  & & & & &     \cr 
 \cline{2-13}
\cr
&\multicolumn{12}{c}{Case D: $\delta= 10^{-5},~{\bf k}=2\pi(2,1)/\sqrt{5}$} \cr
& \multicolumn{6}{l}{\bf density-gas} & \multicolumn{6}{l}{\bf $x-$vel-gas} \cr
   16 &    7.7E-08 &    --  &    1.2E-07 &    --  &    2.4E-07 &    --  &    9.2E-08 &    --  &    1.4E-07 &    --  &    2.9E-07 &    --  \cr 
   32 &    2.1E-08 &    1.8 &    3.4E-08 &    1.8 &    6.8E-08 &    1.8 &    2.6E-08 &    1.8 &    4.1E-08 &    1.8 &    8.2E-08 &    1.9 \cr 
   64 &    5.6E-09 &    1.9 &    8.7E-09 &    1.9 &    1.7E-08 &    2.0 &    6.7E-09 &    1.9 &    1.1E-08 &    1.9 &    2.1E-08 &    1.9 \cr 
  128 &    1.4E-09 &    2.0 &    2.2E-09 &    2.0 &    4.4E-09 &    2.0 &    1.7E-09 &    2.0 &    2.7E-09 &    2.0 &    5.4E-09 &    2.0 \cr 
\cline{2-7} \cline{8-13} \cr
& \multicolumn{6}{l}{\bf $y-$vel-gas} & \multicolumn{6}{l}{\bf pressure} \cr
   16 &    2.3E-08 &    --  &    3.7E-08 &    --  &    7.8E-08 &    --  &  1.0E-07 &    --  &    1.6E-07 &    --  &    3.5E-07 &    --     \cr 
   32 &    5.6E-09 &    2.0 &    8.8E-09 &    2.1 &    1.7E-08 &    2.2 &  3.0E-08 &    1.8 &    4.7E-08 &    1.8 &    9.7E-08 &    1.9   \cr 
   64 &    1.3E-09 &    2.1 &    2.1E-09 &    2.1 &    4.1E-09 &    2.1 &  7.8E-09 &    1.9 &    1.2E-08 &    1.9 &    2.5E-08 &    2.0   \cr 
  128 &    3.2E-10 &    2.1 &    5.0E-10 &    2.1 &    1.0E-09 &    2.0 &  2.0E-09 &    2.0 &    3.1E-09 &    2.0 &    6.2E-09 &    2.0   \cr 
\cline{2-7} \cline{8-13} \cr
& \multicolumn{6}{l}{\bf $x-$magnetic} & \multicolumn{6}{l}{\bf $y-$magnetic} \cr
   16 &    3.9E-08 &    --  &    6.3E-08 &    --  &    1.6E-07 &    --  &    4.7E-08 &    --  &    8.2E-08 &    --  &    1.9E-07 &    -- \cr 
   32 &    9.0E-09 &    2.1 &    1.4E-08 &    2.2 &    3.3E-08 &    2.3 &    1.2E-08 &    2.0 &    1.9E-08 &    2.1 &    4.4E-08 &    2.1\cr 
   64 &    2.1E-09 &    2.1 &    3.3E-09 &    2.1 &    7.3E-09 &    2.2 &    2.9E-09 &    2.0 &    4.6E-09 &    2.1 &    1.0E-08 &    2.1\cr 
  128 &    5.2E-10 &    2.0 &    8.1E-10 &    2.0 &    1.7E-09 &    2.1 &    7.3E-10 &    2.0 &    1.1E-09 &    2.0 &    2.4E-09 &    2.1\cr 
\hline
\hline
\end{tabular*}
\end{small}
\end{table}


\begin{table}  
\begin{small}
\caption{Convergence Rates Slow MHD Waves.\label{slow:tab}}
\begin{tabular*}{\textwidth}{@{\extracolsep{\fill}}lcccccccccccc}
\hline
\hline \cr
N$_{\rm cells}$ & $L_1$ & $R_1$ & $L_2$ & $R_2$ & $L_\infty$ & $R_\infty$ & $L_1$ & $R_1$ & $L_2$ & $R_2$ & $L_\infty$ & $R_\infty$ \cr \cline{2-13} \cr
&\multicolumn{12}{c}{Case E: $\delta= 10^{-5},~{\bf k}=2\pi(1,0)$} \cr
& \multicolumn{6}{l}{\bf density-gas} & \multicolumn{6}{l}{\bf $x-$vel-gas} \cr
   16 &    7.5E-09 &    --  &    1.7E-08 &    --  &    4.7E-08 &    --  &    4.4E-09 &    --  &    9.9E-09 &    --  &    2.8E-08 &    --  \cr 
   32 &    1.9E-09 &    2.0 &    4.2E-09 &    2.0 &    1.2E-08 &    2.0 &    1.1E-09 &    2.0 &    2.5E-09 &    2.0 &    7.0E-09 &    2.0 \cr 
   64 &    4.8E-10 &    2.0 &    1.1E-09 &    2.0 &    3.0E-09 &    2.0 &    2.8E-10 &    2.0 &    6.2E-10 &    2.0 &    1.8E-09 &    2.0 \cr 
  128 &    1.2E-10 &    2.0 &    2.7E-10 &    2.0 &    7.5E-10 &    2.0 &    7.0E-11 &    2.0 &    1.6E-10 &    2.0 &    4.4E-10 &    2.0 \cr 
\cline{2-7} \cline{8-13} \cr
& \multicolumn{6}{l}{\bf $y-$vel-gas} & \multicolumn{6}{l}{\bf pressure} \cr
   16 &    1.4E-08 &    --  &    3.1E-08 &    --  &    8.7E-08 &    --  &  1.1E-08 &    --  &    2.3E-08 &    --  &    6.6E-08 &    --    \cr 
   32 &    3.5E-09 &    2.0 &    7.7E-09 &    2.0 &    2.2E-08 &    2.0 &  2.7E-09 &    2.0 &    5.9E-09 &    2.0 &    1.7E-08 &    2.0   \cr 
   64 &    8.7E-10 &    2.0 &    1.9E-09 &    2.0 &    5.5E-09 &    2.0 &  6.7E-10 &    2.0 &    1.5E-09 &    2.0 &    4.2E-09 &    2.0   \cr 
  128 &    2.2E-10 &    2.0 &    4.8E-10 &    2.0 &    1.4E-09 &    2.0 &  1.7E-10 &    2.0 &    3.7E-10 &    2.0 &    1.1E-09 &    2.0   \cr 
\cline{2-7} \cline{8-13} \cr
& \multicolumn{6}{l}{\bf $y-$magnetic} & \multicolumn{6}{l}{\bf } \cr
   16 &    1.1E-08 &    --  &    2.5E-08 &    --  &    7.1E-08 &    --  &  & & & & &  \cr 
   32 &    2.9E-09 &    2.0 &    6.4E-09 &    2.0 &    1.8E-08 &    2.0 &  & & & & &  \cr 
   64 &    7.2E-10 &    2.0 &    1.6E-09 &    2.0 &    4.5E-09 &    2.0 &  & & & & &  \cr 
  128 &    1.8E-10 &    2.0 &    4.0E-10 &    2.0 &    1.1E-09 &    2.0 &  & & & & &  \cr 
 \cline{2-13}
\cr
&\multicolumn{12}{c}{Case F: $\delta= 10^{-5},~{\bf k}=2\pi(2,1)/\sqrt{5}$} \cr
& \multicolumn{6}{l}{\bf densit$y-$gas} & \multicolumn{6}{l}{\bf $x-$vel-gas} \cr
   16 &    6.1E-08 &    --  &    9.5E-08 &    --  &    1.9E-07 &    --  &    5.7E-08 &    --  &    8.9E-08 &    --  &    1.8E-07 &    --  \cr 
   32 &    1.6E-08 &    2.0 &    2.4E-08 &    2.0 &    4.9E-08 &    2.0 &    1.4E-08 &    2.0 &    2.3E-08 &    2.0 &    4.7E-08 &    1.9 \cr 
   64 &    3.9E-09 &    2.0 &    6.2E-09 &    2.0 &    1.2E-08 &    2.0 &    3.7E-09 &    2.0 &    5.8E-09 &    2.0 &    1.2E-08 &    2.0 \cr 
  128 &    9.9E-10 &    2.0 &    1.6E-09 &    2.0 &    3.1E-09 &    2.0 &    9.2E-10 &    2.0 &    1.4E-09 &    2.0 &    2.9E-09 &    2.0 \cr 
\cline{2-7} \cline{8-13} \cr
& \multicolumn{6}{l}{\bf $y-$vel-gas} & \multicolumn{6}{l}{\bf pressure} \cr
   16 &    1.4E-07 &    --  &    2.2E-07 &    --  &    4.4E-07 &    --  &   8.5E-08 &    --  &    1.3E-07 &    --  &    2.8E-07 &    --   \cr 
   32 &    3.6E-08 &    2.0 &    5.7E-08 &    2.0 &    1.1E-07 &    1.9 &   2.2E-08 &    2.0 &    3.5E-08 &    2.0 &    7.0E-08 &    2.0   \cr 
   64 &    9.1E-09 &    2.0 &    1.4E-08 &    2.0 &    2.9E-08 &    2.0 &   5.6E-09 &    2.0 &    8.7E-09 &    2.0 &    1.8E-08 &    2.0   \cr 
  128 &    2.3E-09 &    2.0 &    3.6E-09 &    2.0 &    7.2E-09 &    2.0 &   1.4E-09 &    2.0 &    2.2E-09 &    2.0 &    4.4E-09 &    2.0   \cr 
\cline{2-7} \cline{8-13} \cr
& \multicolumn{6}{l}{\bf $x-$magnetic} & \multicolumn{6}{l}{\bf $y-$magnetic} \cr
   16 &    6.1E-08 &    --  &    9.6E-08 &    --  &    2.0E-07 &    --  &    6.3E-08 &    --  &    9.9E-08 &    --  &    2.2E-07 &    --  \cr 
   32 &    1.5E-08 &    2.0 &    2.4E-08 &    2.0 &    5.0E-08 &    2.0 &    1.6E-08 &    1.9 &    2.6E-08 &    1.9 &    5.5E-08 &    2.0 \cr 
   64 &    3.9E-09 &    2.0 &    6.1E-09 &    2.0 &    1.2E-08 &    2.0 &    4.1E-09 &    2.0 &    6.5E-09 &    2.0 &    1.3E-08 &    2.0 \cr 
  128 &    9.8E-10 &    2.0 &    1.5E-09 &    2.0 &    3.1E-09 &    2.0 &    1.0E-09 &    2.0 &    1.6E-09 &    2.0 &    3.3E-09 &    2.0 \cr 
\hline
\hline
\end{tabular*}
\end{small}
\end{table}

\subsection{Riemann Problem} \label{riemann:se}
\begin{table}  [ht]
\caption{Riemann Problem Set\label{riemannset:tab}}
\begin{tabular*}{\textwidth}{@{\extracolsep{\fill}}lccccccccccccccc}
\hline \hline
& \multicolumn{7}{c}{\bf left-state} && \multicolumn{7}{c}{\bf right-state}
\cr 
Test & $\rho$ & $u_{x}$ & $u_{y}$ & $u_{z}$ & $p$ & $B_{y}$ &  $B_{z}$ && $\rho$ & $u_{x}$ & $u_{y}$ & $u_{z}$ & $p$ & $B_{y}$ &  $B_{z}$   \cr
\cline{0-1} \cline{2-8}  \cline{10-16}
{\tiny Brio-Wu}& 1 & 0 & 0 & 0 & 1 & 1 & 0 && 0.128 & 0 & 0 & 0 & 0.1 &-1 & 0  \cr
{\tiny Dai-Woodward}&1.08 &1.2 &0.01 &0.5 &0.95 &$\frac{3.6}{\sqrt{4\pi}}$ &$\frac{2}{\sqrt{4\pi}}$ && 1& 0& 0& 0& 1& $\frac{4}{\sqrt{4\pi}}$& $\frac{2}{\sqrt{4\pi}}$ \cr
{\tiny Ryu-Jones}& 1 & 0 & 0& 0& 1& 0& 0&& 0.3& 0& 0& 1& 0.2& 1&0  \cr
{\tiny Fast Raref.}& 1& -2& 0& 0& 0.45& 0.5& 0&& 1& 2& 0& 0& 0.45& 0.5& 0 \cr
\hline
\end{tabular*}
\end{table}
We next consider a set of Riemann problems from the literature which
are standard tests for MHD algorithms. The Riemann problem is described
in general by the following initial-value problem
\begin{equation} \label{ic:eq}
W [x_0,t=0] = 
   \left\{ \begin{array}{lll}  
   W_{left} & \mbox{if} &  x_0\le 0.5\\ 
   W_{right} &  \mbox{if} &  x_0>0.5
    \end{array} \right. 
\end{equation}
where $W_{left/right}$ represents the primitive variables to the
left/right of the initial discontinuity. The set of problems and
corresponding initial conditions are summarized in
Table~\ref{riemannset:tab}, except for the value of the $x-$component
of the magnetic field which will be specified for each problem
explicitly. In all cases we use a CFL number $C_{\rm CFL}=0.8$, third
order PPM reconstruction scheme and characteristic limiting.  The
domain size is $L=1$ and the boundary conditions are simply,
$W(x_0=0,t)=W_{left}$, and, $W(x_0=1,t)=W_{right}$.

We should point out
that~\citet{MiyoshiKusano2005JCoPh.208..315M} have carried out
extensive tests of their HLLD solver, which we employ in our code,
particularly to compare its performance with that of Roe and other
solvers of the HLLC family.  We shall repeat some of those tests
here. Note that while in most
cases~\citet{MiyoshiKusano2005JCoPh.208..315M} used a first order
accurate piece-wise constant reconstruction scheme, we have
used the third order accurate PPM method, 
so the solution profiles in our plots appear sharper.  The
tests which we will perform below involve the full set of MHD waves,
including those associated with the slow mode which is not included
explicitly in the HLLD solver. However, in accord 
with~\citet{MiyoshiKusano2005JCoPh.208..315M},
we find that in these tests 
the full MHD structure of the solution, including features
associated with the slow mode, is correctly reproduced.

\subsubsection{Brio \& Wu}
\begin{figure*}
\plotone{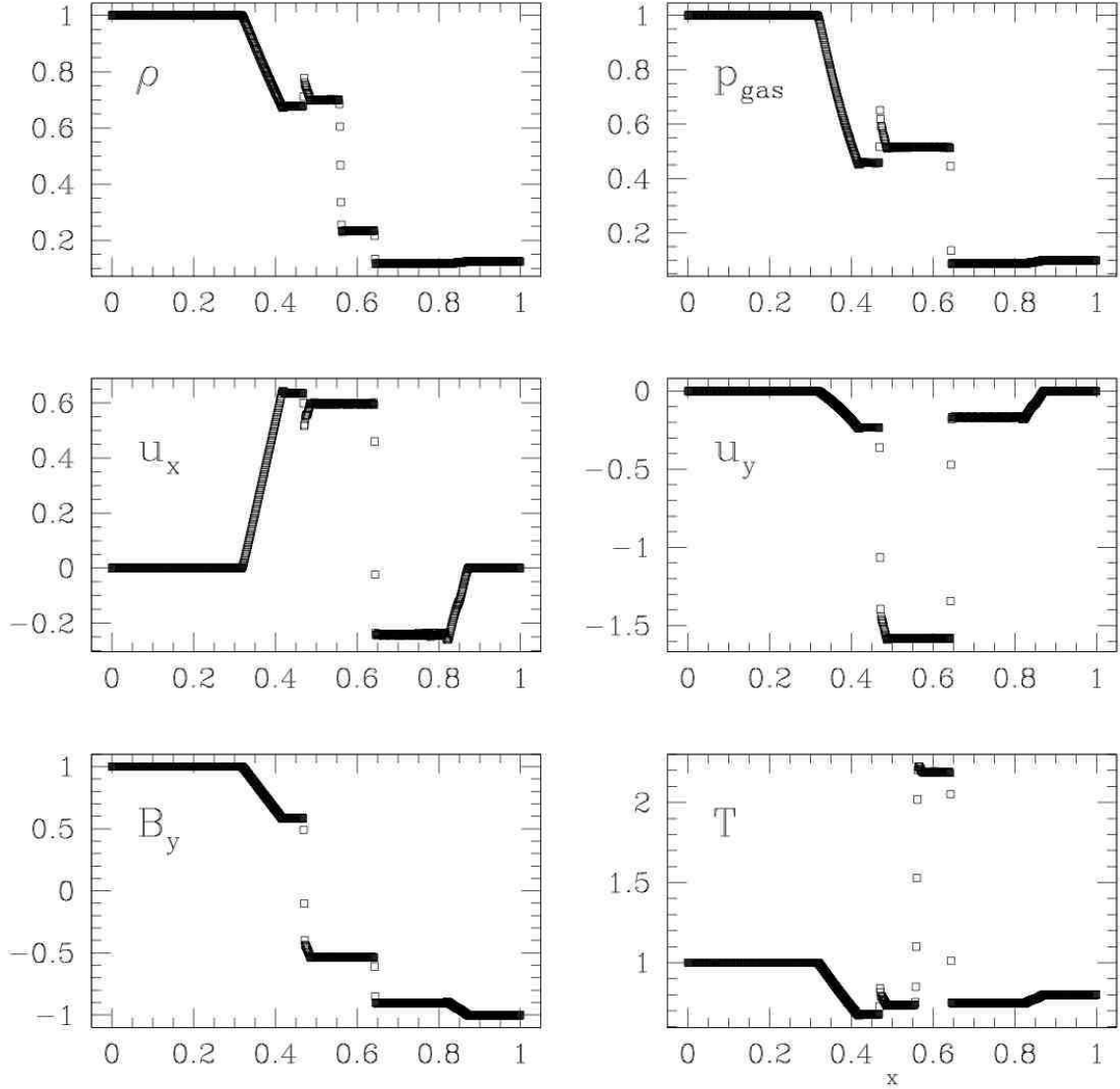} \\
\caption{Brio \& Wu shock tube problem: solution for $t=0.1$ solved on
  a grid using 800 zones. See Table~\ref{riemannset:tab} for the
  initial conditions.  From left to right and top to bottom shown
  are, respectively: density, pressure, velocity components along the
  x and y axis, magnetic field component along the $y-$axis and
  temperature.}
\label{briowu:fig}
\end{figure*}
We begin with the Riemann problem presented
in~\cite{BrioWu1988JCoPh..75..400B}, listed at the top of
Table~\ref{riemannset:tab}. The $x-$component of the magnetic field is
$B_x=0.75$ and the adiabatic index $\gamma=2$ as in the original
paper. Following common practice, the problem is solved on a grid with
800 points along the $x-$axis and the solution is computed until time
$t=0.1$.  The results are shown in Fig.~\ref{briowu:fig}.  All the
solution features are well reproduced, including, from left to right,
a fast rarefaction followed by a slow compound wave, moving to the
left, and a contact discontinuity, slow shock and fast rarefaction
moving to the right~\citep{BrioWu1988JCoPh..75..400B}.  As usual with
shock-capturing methods, shocks are resolved with a couple of zones
throughout the duration of the calculation, while the contact
discontinuities, captured here within a few zones, tend to spread out
over time.  There are some oscillations in the $x-$component of the
velocity field.  These are not present if we adopt a PLM
reconstruction scheme, but worsen if we switch from a characteristic
to a primitive limiting scheme.

\subsubsection{Dai \& Woodward} \label{daiwoodward:se}
\begin{figure*}
\plotone{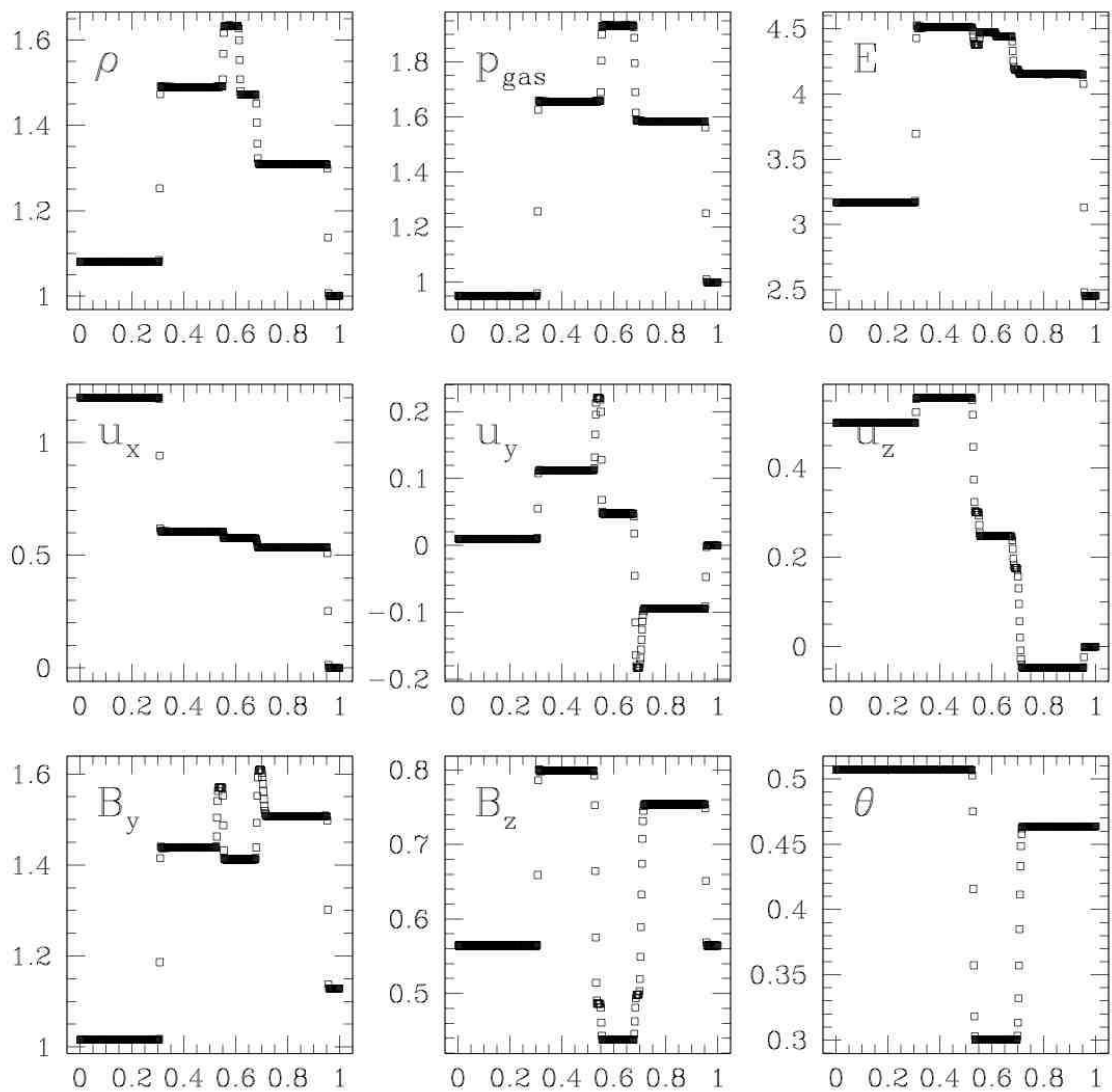} \\
\caption{Dai \& Woodward shock tube problem: solution for $t=0.2$
  solved on a grid using 512 zones. See Table~\ref{riemannset:tab} for
  the initial conditions.  From left to right and top to bottom, shown
  are, respectively: density, pressure, energy, three velocity
  components magnetic field components, along the y and z axis, and
  $\theta=\tan^{-1}(B_z/By)$.}
\label{daiwoodward:fig}
\end{figure*}
The second Riemann problem listed in table~\ref{riemannset:tab} is
taken from~\cite{DaiWoodward1998ApJ...494..317D}.  In this case the
$x-$component of the magnetic field is $B_x=4/\sqrt{4\pi}$ and the
adiabatic index $\gamma=5/3$. The problem is solved on a grid with 512
points along the $x-$axis and the solution is computed until time
$t=0.2$.  The results are shown in Fig.~\ref{daiwoodward:fig}.  The
initial conditions for this problem expose the full eigenstructure of
the MHD system, as they produce three pairs of MHD waves traveling in
opposite directions with respect to the initial discontinuity, in
addition to the contact wave. The waves include fast and slow shocks
responsible, among others, for the jumps in the pressure and velocity
fields, the contact wave which appears in the density field alone, and
the rotational discontinuity which affects the magnetic field
components alone.  As in the previous case, while all discontinuities
are well reproduced, shocks are the sharpest features captured with
about two zones.

\subsubsection{Ryu \& Jones}
\begin{figure*}
\plotone{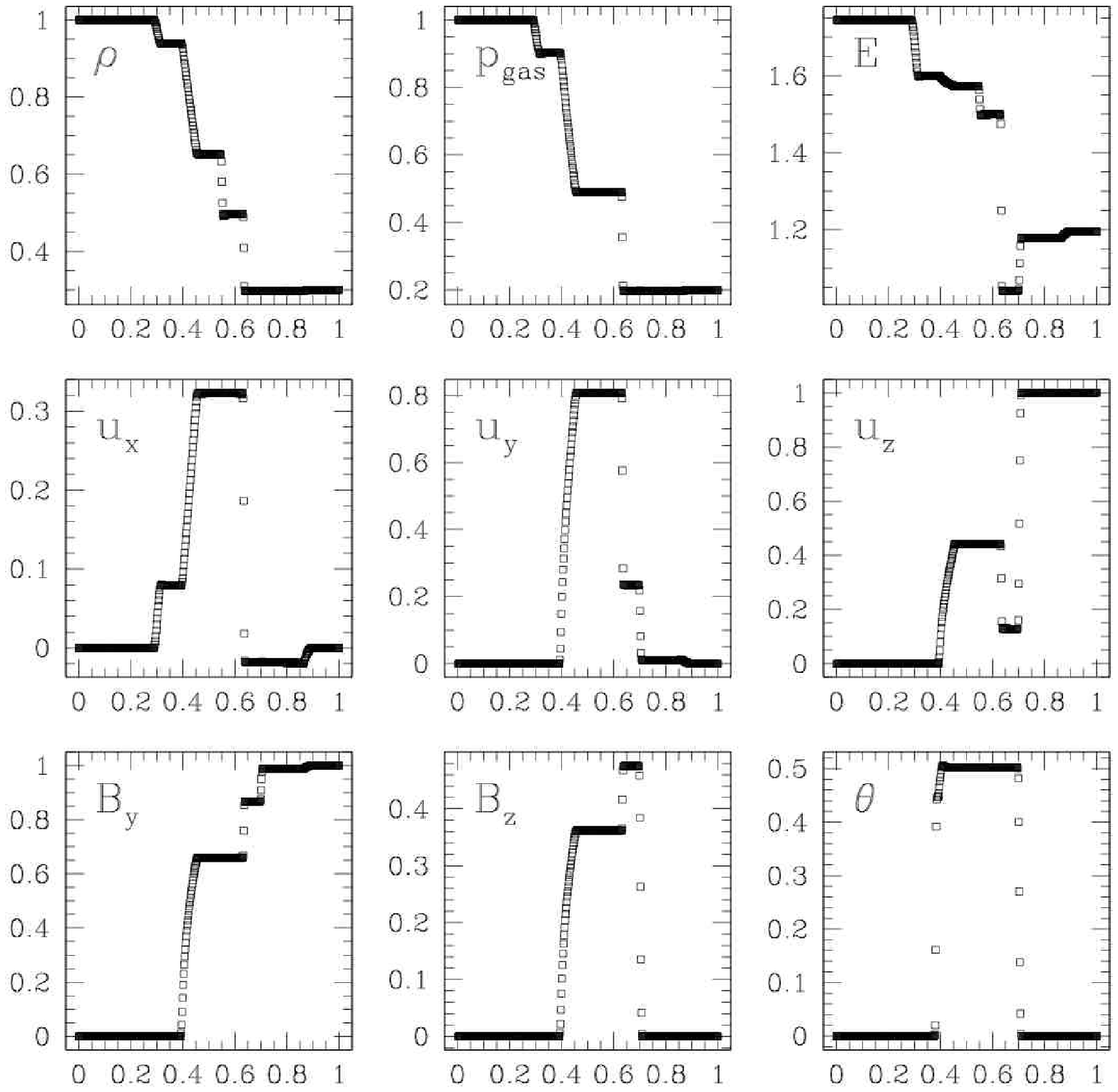} \\
\caption{Ryu \& Jones shock tube problem: solution for $t=0.16$
  solved on a grid using 512 zones. See Table~\ref{riemannset:tab} for
  the initial conditions.  From left to right and top to bottom, shown
  are, respectively: density, pressure, energy, three velocity
  components, magnetic field components along the y and z axis, and
  $\theta=\tan^{-1}(B_z/By)$.}
\label{ryujones:fig}
\end{figure*}
The third Riemann problem in our table~\ref{riemannset:tab} is one of
the many problems studied by~\cite{RYUJONES1995ApJ...442..228R}.  Here
$B_x=0.7$ and $\gamma=5/3$.  The problem is solved on a grid with 512
points along the $x-$axis and the solution is computed until time
$t=0.16$, as in the original paper.  The results are shown in
Fig.~\ref{ryujones:fig}.  The solution to this problem includes, from
left to right, a hydrodynamic rarefaction, a switch-on slow shock, a
contact discontinuity, a slow shock, a rotational discontinuity and a
fast rarefaction.  Switch-on/off fast/slow shocks are interesting MHD
solutions that cause the tangential magnetic field to turn on/off,
respectively.  As in the previous tests the structure of the solution
is well reproduced, including the features associated with the slow
MHD mode, and the discontinuities are captured within a few to several
zones.

\subsubsection{Fast rarefaction}

\begin{figure*}
\plotone{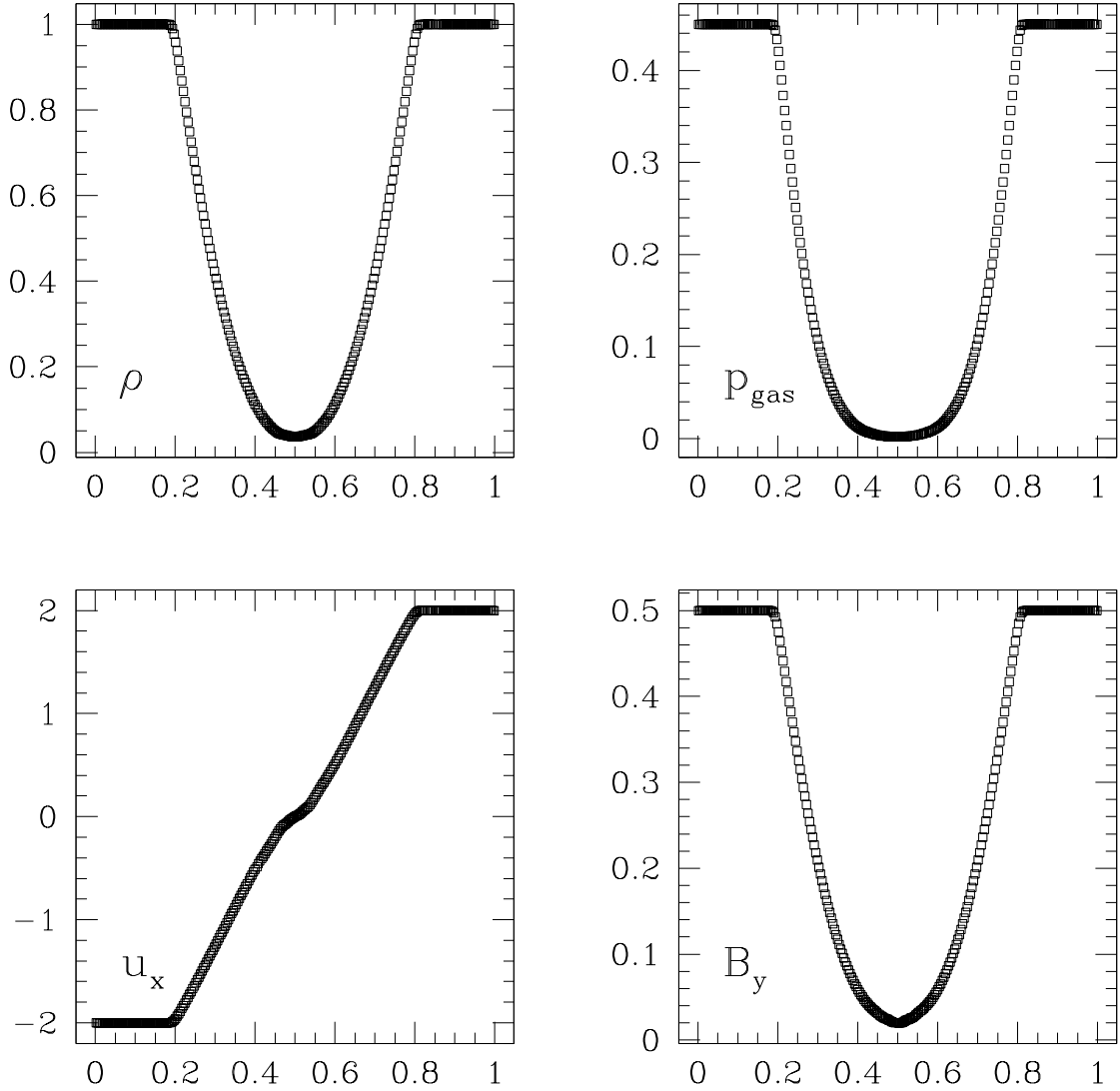} \\
\caption{Fast rarefaction: solution for $t=0.16$
  solved on a grid using 512 zones. See Table~\ref{riemannset:tab} for
  the initial conditions.  From left to right and top to bottom, shown
  are, respectively: density, gas pressure, $x-$component of the velocity
and $y-$component of the magnetic field.}
\label{strongraref:fig}
\end{figure*}

The last one-dimensional Riemann problem listed in
table~\ref{riemannset:tab}  is similar to the ones
presented in~\cite{Einfeldt1991JCoPh..92..273E}.  Its purpose is to
test the method/code robustness in the case of flows in which
the energy is dominated by the kinetic component and unphysical states
with negative density or internal energy can arise.  This problem is
solved using a grid with 256 grid points. The solution at time $t=0.1$
is shown in Fig.~\ref{strongraref:fig}. 
The results in Fig.~\ref{strongraref:fig} show a stable solution
which correctly reproduces the two rarefaction waves
propagating away from the grid midpoint. 
A very similar test has also been
performed by~\citet{Stone2008ApJS..178..137S}, with similar results,
and~\citet{MiyoshiKusano2005JCoPh.208..315M}.
The latter authors actually use a faster expansion velocity,
namely $u_{x,\frac{L}{R}}=\mp 3$, in their initial conditions and show that
when used with a (first order) Godunov's method
the numerical solution remains stable. While we are able to 
reproduce their result we find that at high resolution
some spurious oscillations can be generated when a higher-order
Godunov's method is used. This may suggest the need for artificial viscosity
in the case of highly supersonic expansions with higher-order
Godunov's methods.
\subsubsection{Inclined Dai \& Woodward Shock-Tube Problem}
\begin{figure*}
\plotone{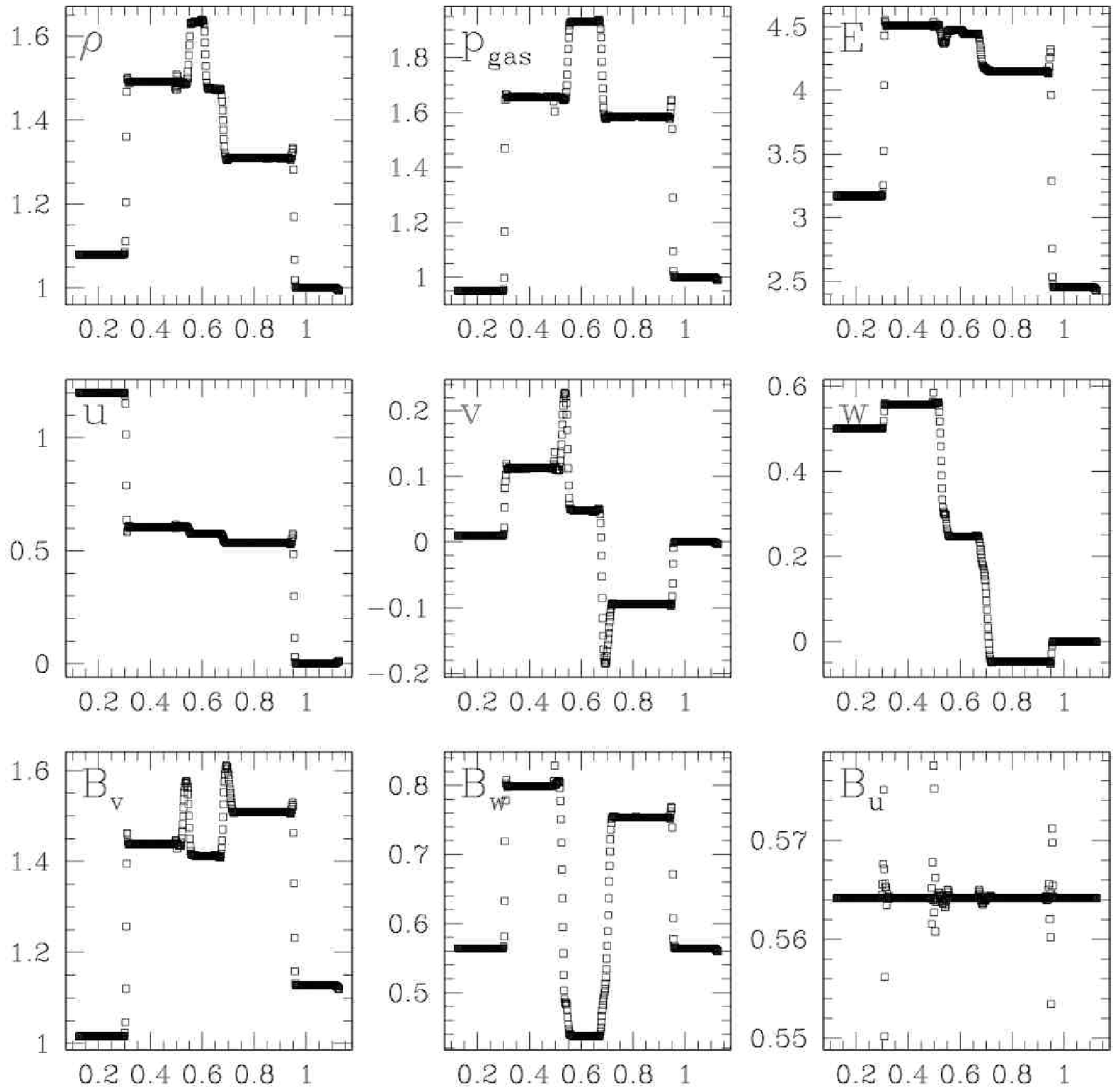} \\
\caption{Inclined version of Dai \& Woodward shock tube problem:
  solution for $t=0.2$ solved on a grid using 512 zones. See
  Table~\ref{riemannset:tab} for the initial conditions.  From left to
  right and top to bottom, shown are, respectively: density, pressure,
  energy, three velocity components and three magnetic field
  components.}
\label{daiwoodward2d:fig}
\end{figure*}

We have repeated the problem in Sec.~\ref{daiwoodward:se} with the
initial discontinuity inclined with respect to the grid such that
its normal has components $\vec n=(2,1)/\sqrt{5}$.  This problem
tests the ability of the code to reproduce one-dimensional solutions
when they are not aligned with the grid.  Besides the numerical tests,
there are a few other complications related with the numerical set-up
of this problem. First, the boundary conditions need revision. To
avoid the implementation of ``shift-periodic'' boundary
conditions~\citep{Toth2000JCoPh.161..605T}, we use a domain with size
$(2L,L)\,2/\sqrt{5}$. This allows us to accommodate two adjacent
identical problems along the direction, $\vec n=(2,1)/\sqrt{5}$, and
apply periodic boundary conditions to the domain.~\footnote{Obviously,
  since the sequence of states will be $W_{left},W_{right},
  W_{left},W_{right}$, the interaction at the second interface will
  produce a similar Riemann problem with inverted initial
  conditions. However, we will present only part of the solution,
  selected for proper comparison with the analogous one-dimensional
  problem in Sec.~\ref{daiwoodward:se}.}  In addition, since the magnetic field rotates across the
discontinuity, the initialization is non-trivial and, unless a special
precaution is taken, e.g. by deriving the magnetic field from a vector
potential, the initial magnetic field will not be divergence-free
(this is not an issue in the special case in which $\vec n$ is along
the diagonal and there is symmetry among the coordinate axis).  In our
case no such precaution is taken and we just remap the initial
conditions onto the rotated grid.  As a result there is a jump in the
normal component of the magnetic field across the discontinuity. This
causes some minor artifacts with respect to the one-dimensional
solution. This is acceptable since we are interested in making sure
that the structure of the one-dimensional solution is reproduced with
fidelity and the waves propagate at the correct speed.

In order to solve the problem, we have covered the domain with a grid
of $1144\times 572$ cells. This corresponds to 512 cells
along the direction perpendicular to $\vec n$, which is equivalent to
the resolution used in Sec.~\ref{daiwoodward:se}.
The results are shown in Fig.~\ref{daiwoodward2d:fig} where we plot 
the values of the solution along the first row of the computational 
domain, starting from $\half L+\delta_L$ to $\frac{3}{2}L+\delta_L$.
The starting point is shifted by $\delta_L\sim 0.2$ to the left,
to hide the perturbation of
$W_{left}$ due to the interaction with the state to its left (which
is $W_{right}$). The vectorial components $(u,v,w)$ are the equivalent
of $(x,y,z)$ in the rotated system.

Fig.~\ref{daiwoodward2d:fig} shows that the one-dimensional solution
is correctly recovered when the plane of the discontinuity is inclined
with respect to the grid.  We notice some oscillation at the fast
magnetosonic shocks, which to some extent are also present in Fig. 7
of~\cite{GardinerStone2008JCoPh.227.4123G}.  These are probably
associated with the oscillations in the normal component of the
magnetic field, also reported in the last panel of
Fig.~\ref{daiwoodward2d:fig}.  These can perhaps be suppressed with
more aggressive limiters~\citep{LondrillodelZanna2004JCoPh.195...17L}.
There is also a spurious feature, ahead of the left-moving rotational
discontinuity, which is probably due to the non-solenoidal character
of the initial magnetic field.  However, none of these features
affect either the jump conditions or the wave speeds, as
attested to by the very good correspondence between the solution in
Fig.~\ref{daiwoodward2d:fig} and the one-dimensional counterpart in
Fig.~\ref{daiwoodward:fig}.

\subsection{Magnetic Loop Advection}

In this section we test the ability of the code to follow the
advection of a loop of weak magnetic field frozen in a background flow.
This problem is non-trivial for conservative schemes 
and~\cite{GardinerStone2005JCoPh.205..509G,GardinerStone2008JCoPh.227.4123G}
emphasize that spurious results will be produced as a result
of improper account of the MHD source terms entering the 
predictor step, as discussed in Sec.~\ref{nps:se}.
The initial conditions are detailed
in~\cite{GardinerStone2008JCoPh.227.4123G}.  The loop is basically a
tube of magnetic flux frozen in a 
medium with unit density and pressure, $\rho=p_{gas}=1$
and uniform advection velocity, ${\bf u}_{loop}$, to be specified below.

We carry out the test both in a 2D or a 3D geometry.
For the 2D case, we align the loop axis with the $x_2$
coordinate axis of the computational domain.
The vector potential from which the magnetic field
is initialized is then given by, ${\bf A}=(0,0,A_2)$, with
\begin{equation} \label{vectpot:eq}
A_2=
   \left\{ \begin{array}{lll}  
   B_0(R-r) & \mbox{if} &  r\le R,\\
    0 &  \mbox{if} &  r>R,
    \end{array} \right. 
\end{equation}
where $B_0=10^{-3}$, $r=\sqrt{x_0^2+x_1^2}$, and, $R=0.3$, is the
radius of the tube.
The computational domain itself consists of a rectangular box of 
dimensions $(2N,N)$ with periodic boundaries, 
and the loop advection velocity is ${\bf u}_{loop}=(2,1)$.

For the 3D configuration, the axis of the tube is tilted around the $x_1$
axis by $\theta=\arctan 2$ radians (clockwise) and the tube is advected
with velocity ${\bf u}_{loop}=(1,1,2)$.  The domain is still periodic
but has dimensions $(N,N,2N)$.  The vector potential is still defined
as in~(\ref{vectpot:eq}) but with respect to the new coordinates
\begin{eqnarray} \label{trasf:eq} \notag
x^\prime_0 &=& x_0\cos\theta +x_2\sin\theta , \\
x^\prime_1 &=& x_1, \\
x^\prime_2 &=& -x_0\sin\theta +x_2\cos\theta. \notag
\end{eqnarray}

\begin{figure*}
\plotone{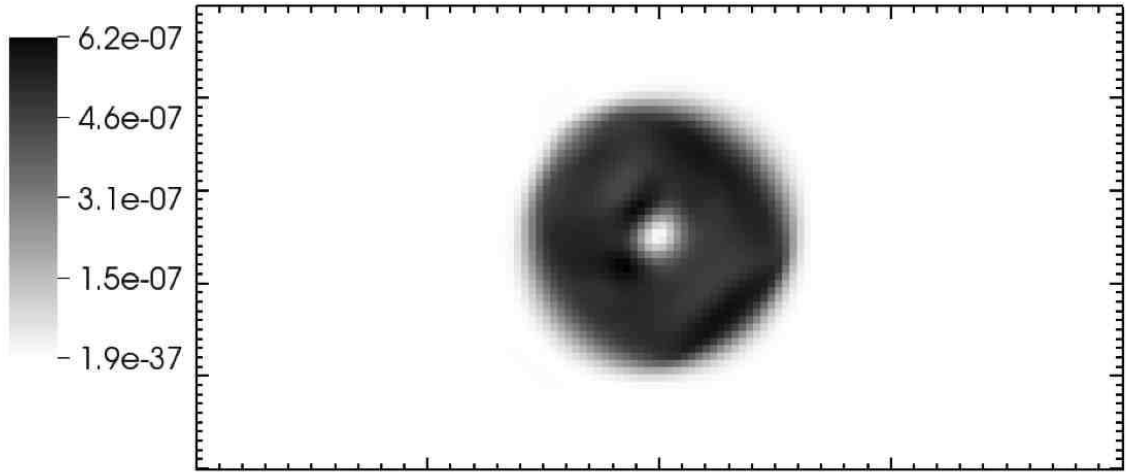} \\
\caption{Loop advection in 2D: magnetic pressure at $t=2$ from a
  calculation using $N=64$, and corresponding colorbar (top left).}
\label{loop2D:fig}
\end{figure*}

\begin{figure*}
\epsscale{.5}
\plotone{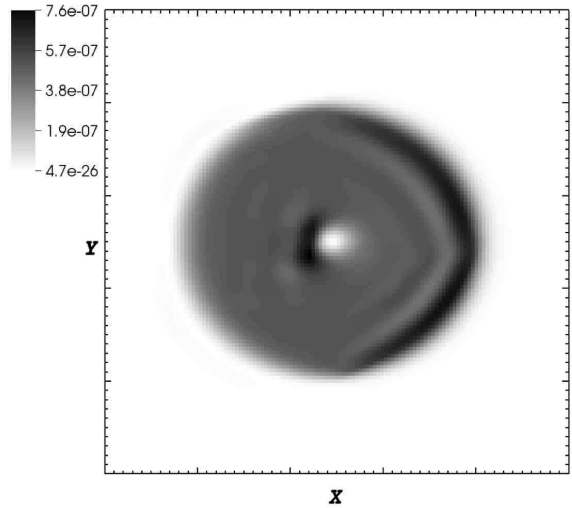} \\
\caption{Loop advection in 3D: cut of the magnetic pressure at $z=0.5$
  of the magnetic pressure at $t=2$ from a calculation using $N=128$,
  and corresponding colorbar (top left).}
\label{loop3D:fig}
\end{figure*}

\begin{figure*}
\plotone{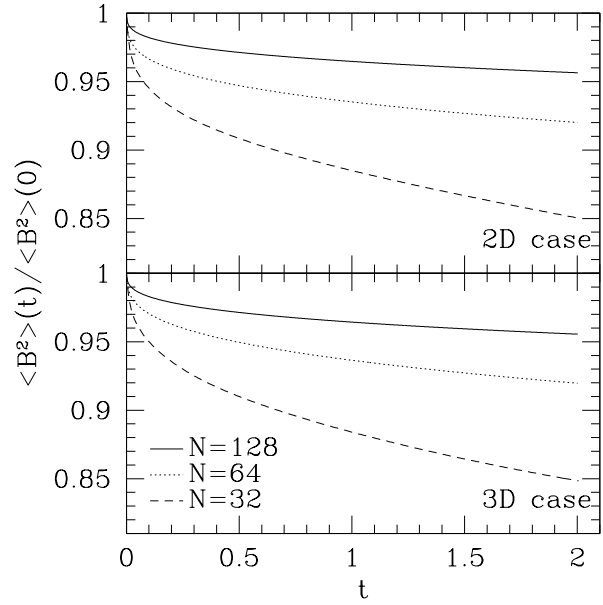} \\
\caption{Loop advection: time evolution of the magnetic energy for the
  2D (top) and 3D (bottom) cases, for three different resolutions.}
\label{ebloop:fig}
\end{figure*}

\begin{figure*}
\plotone{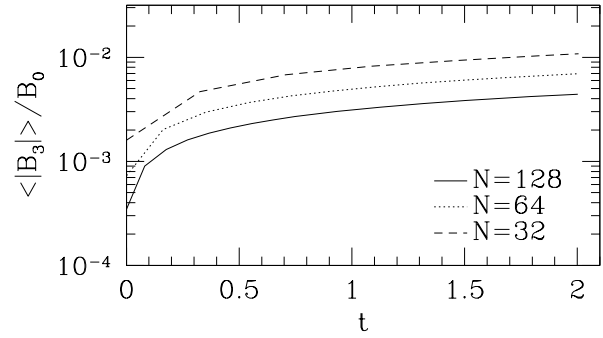} \\
\caption{Loop advection in 3D: time evolution of $B_3$, the magnetic
  field component aligned with the loop axis, for three different
  resolutions.}
\label{b3:fig}
\end{figure*}

The results of this test are first illustrated in
Figures~\ref{loop2D:fig} and~\ref{loop3D:fig} which show the magnetic
pressure at $t=2$ for the 2D and 3D cases respectively.  In the former
case $N=64$, while in the latter case $N=128$.  The results are
comparable to other MHD implementations, in
particular~\citet{GardinerStone2005JCoPh.205..509G,GardinerStone2008JCoPh.227.4123G}. As
pointed out by those authors, the magnetic pressure suffers
dissipation mostly close to the loop center (where the curl of $\vec
B$ is singular) and boundary. However, both in the 2D and 3D cases,
the loop retains to a good extent its initial symmetry and energy.
This latter point is further illustrated in Fig.~\ref{ebloop:fig},
reporting the evolution of the normalized magnetic pressure up to
$t=2$ for three different resolutions.  One important aspect of the 3D
version of this test is to check the ability of the scheme to keep the
magnetic field component along the loop axis, $B_3$, close to
zero. This is important here because we have not strictly followed the
recommendation of~\citet{GardinerStone2008JCoPh.227.4123G} when
implementing the MHD source terms in the predictor step. The evolution
of $B_3$ is reported in Figure~\ref{b3:fig}, again for three different
values of the resolution, up to $t=2$. Due to the simpler form of the
employed MHD source terms, $\langle|B_3|\rangle$ is larger than found
in~\citet{GardinerStone2008JCoPh.227.4123G} at the same time ($t=1$),
but only slightly so and $B_3$
remains negligible compared to the total magnetic field.


\subsection{Orszag-Tang Vortex}

\begin{figure*}
\plotone{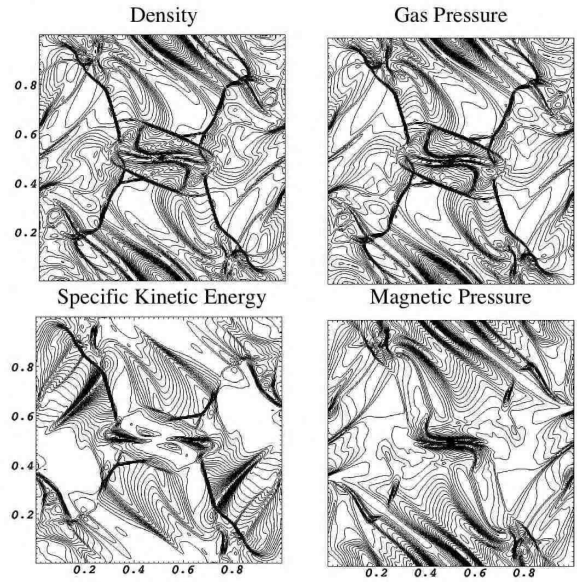} \\
\caption{Orszag-Tang Vortex: thirty equally-spaced contour levels
  between max and min value of the numerical solution at $t=0.5$
  respectively for density (top-left), gas pressure (top-right),
  specific kinetic energy (bottom-left) and magnetic pressure
  (bottom-right).}
\label{vortex:fig}
\end{figure*}

\begin{figure*}
\plotone{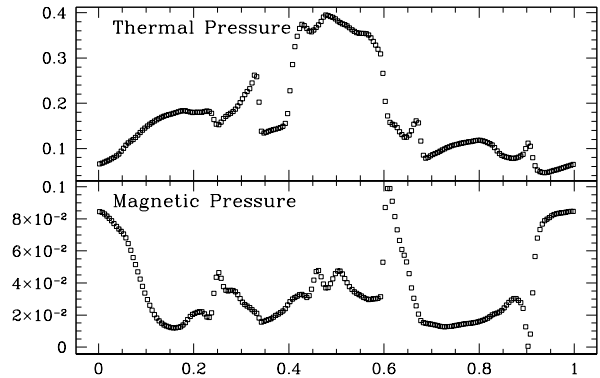} \\
\caption{Orszag-Tang Vortex: One-dimensional cuts along the $x-$axis at
  $y=0.4277$ for the thermal pressure (top) and magnetic pressure
  (bottom).}
\label{vortex_slice:fig}
\end{figure*}

We now turn to a series of classical multidimensional tests for MHD
codes.  First, we compute the compressible Orszag-Tang vortex
problem~\citep{OrszagTang1979JFM....90..129O}.  We solve this problem
on a computational domain of size $L=1$ with periodic boundary
conditions and a rectangular grid of 200$\times$200 cells. We use an
adiabatic index $\gamma=5/3$.  We present results obtained using the
PPM reconstruction scheme and primitive limiting, but very similar
results are produced using characteristic limiting.  The initial
conditions in terms of the primitive variables are as follows
\begin{equation}
W =\left[\rho_0,-u_0\sin(2\pi y),u_0 \sin(2\pi x),0,P_0,-B_0\sin(2\pi y),
B_0 \sin(4\pi x),0\right]^T
\end{equation}
where $\rho_0=25/36\pi$, $P_0=5/12\pi$ and $u_0$ and $B_0$ are defined
in terms of the sonic Mach number, ${\cal M}=1$, and plasma beta,
$\beta=10/3$, respectively.  Although the initial conditions are
smooth, eventually the flow develops a complex structure with sharp
features and discontinuities. In Fig.~\ref{vortex:fig} we show the
contours of the numerical solution for density, gas pressure, kinetic
energy and magnetic pressure at time $t=0.5$. One dimensional cuts
along the line $y=0.4277$ for the thermal pressure (top) and magnetic
pressure (bottom) are also shown in Fig.~\ref{vortex_slice:fig}.  The
code maintains the symmetry of the solution with respect to the
central point. In addition, Fig.~\ref{vortex_slice:fig} shows that
discontinuous features are captured within a few zones. Finally, the
solution can be compared with similar plots at the same solution time
produced by other
authors~\citep[e.g.][]{rmjf98,Toth2000JCoPh.161..605T,Stone2008ApJS..178..137S},
from which it appears that the code produces the correct flow
structures.


\subsection{Rotor Problem}

\begin{figure*}
\plotone{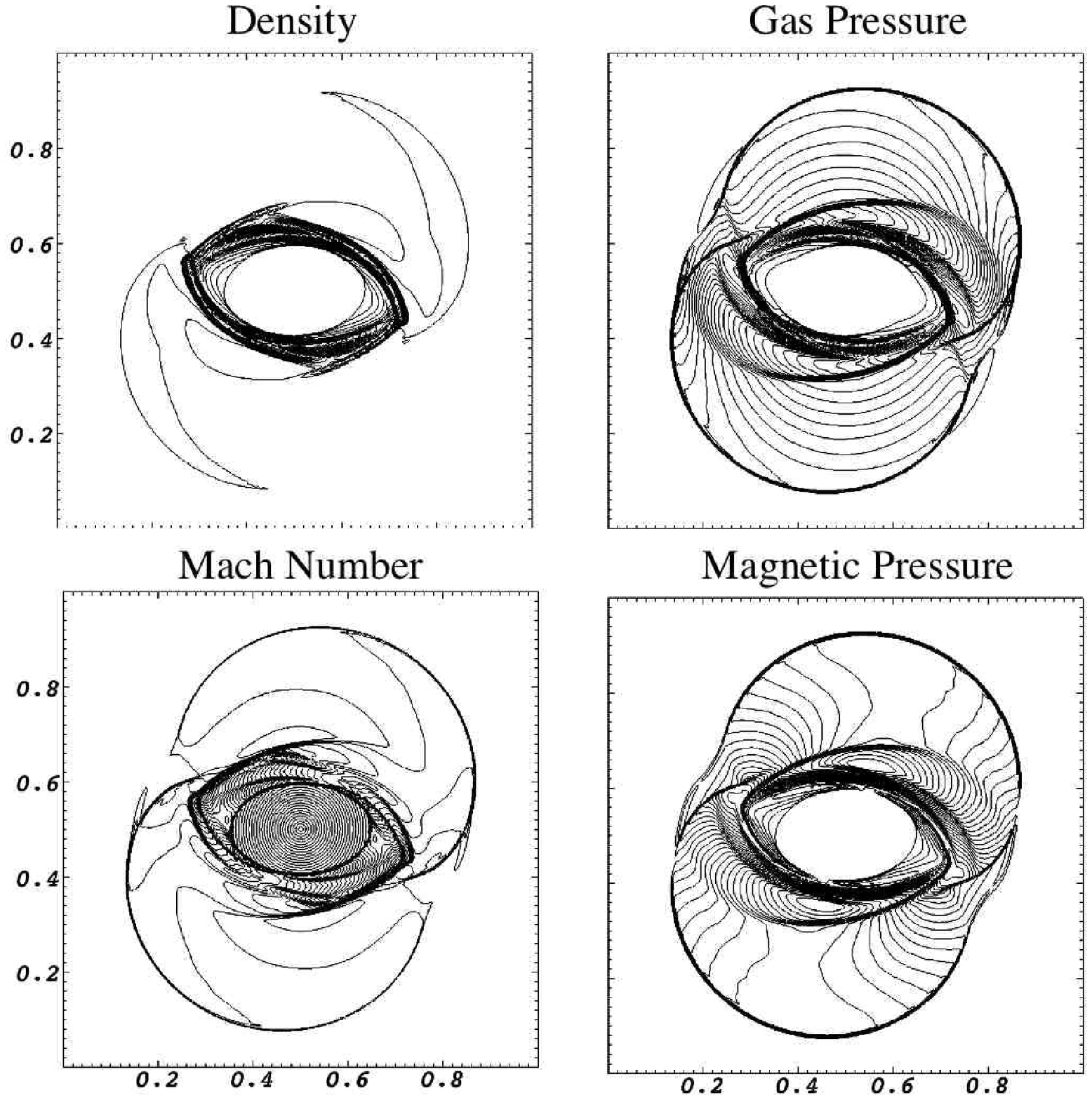} \\
\caption{Rotor Problem: thirty equally-spaced contour levels
  between max and min value of the numerical solution at $t=0.15$
  respectively for density (top-left), gas pressure (top-right),
  Mach number (bottom-left) and magnetic pressure (bottom-right).}
\label{rotor:fig}
\end{figure*}

Another two-dimensional problem commonly used as a test for
multidimensional MHD codes is the {\it rotor} problem described
in~\cite{BalsaraSpicer1999JCoPh.149..270B}.  It consists of a rotating
disk of dense material, threaded by a magnetic field initially
directed along the $x-$axis, and embedded in a tenuous ambient medium
at rest.  As the rotor spins, it winds up the magnetic field lines,
generating Alfv\'en waves which propagate into the ambient medium.
The problem is solved on a computational domain of size $L=1$
with periodic boundary conditions and covered with a rectangular grid
of 400$\times$400 cells. The adiabatic index is $\gamma=1.4$. We use
the PPM reconstruction scheme and primitive limiting, although the use
of characteristic limiting produces very similar results. The
setup of the initial conditions corresponds to the first rotor problem
discussed in~\cite{Toth2000JCoPh.161..605T}, i.e.
\begin{equation} \label{rpic:eq}
W [x,t=0] = 
   \left\{ \begin{array}{lll}  
   W_{disk} & \mbox{if} &  r< r_{disk},\\ 
   W_{amb} &  \mbox{if} &  r> r_{amb},
    \end{array} \right.
\end{equation}
where
$W_{disk}=[\rho_{disk},-\frac{u_0}{r_0}(y-0.5),\frac{u_0}{r_0}(x-0.5),0,
P_0,B_0,0,0]^T$,
$W_{amb}=(\rho_{0},0,0,0,P_0,B_0,0,0)^T$,
$r\equiv\sqrt{x^2+y^2}$, $r_{disk}$ defines the disk
radius and $r_{amb}$ delimits the ambient medium.
In the transition region between the ambient medium and the rotor
the density and velocity fields are interpolated according to
$\rho=(\rho_{disk}-\rho_0)f(r)+\rho_0$,
$u_x=-f(r)u_0(y-0.5)/r$ and $u_y=f(r)u_0(x-0.5)/r$,
with $f(r)=(r_{amb}-r)/(r_{amb}-r_{disk})$,
whereas density and pressure remain uniform.
The results for this test are presented in Fig.~\ref{rotor:fig},
which effectively reproduces Fig. 18 of~\cite{Toth2000JCoPh.161..605T}.
The numerical solution appears very well behaved and the code seem
to pass this test as well.

\section{Extension to  Adaptive Mesh Refinement} \label{amr::se}

Following \cite{bergercolella89} and
\citet{Balsara2001JCoPh.174..614B}, we employ block-structured local
refinement to increase the computational resolution where the accuracy of
the solution needs to be improved. Our implementation is basically an
extension of the MHD case of~\cite{mico07b}.

Generalizing the case discussed in Sec.~\ref{dvo:se}, the
problem domain is discretized on a hierarchy of grids,
$\Gamma^0\dots\Gamma^{\ell_{max}}$, each with its own spacing $h^\ell$ and
refinement ratio $n^\ell_{ref}\equiv h^\ell/h^{\ell+1}$. We assume that
refinement ratio is always even.  

Calculations are performed on a
hierarchy of meshes $\{\Omega^\ell\}_{\ell=0}^{\ell=\ell_{max}}$ such that for
each $\ell$, $\Omega^\ell\subset\Gamma^\ell$.
The base-level uniform rectangular mesh
spans the domain, so $\Omega^0=\Gamma^0$.
Cells for which improved resolution
is desired are marked for refinement, grouped together into logically
rectangular regions, and refined by a factor of $n_{ref}^0$ 
to create the level 1 domain
$\Omega^1$. Further refinement levels, $\Omega^{\ell}$, may then be
created as needed in the same way starting from a refinement level
$\Omega^{\ell-1}$, with refinement ratio $n_{ref}^{\ell-1}$.
The set of generated meshes $\Omega^\ell$ is assumed to be properly nested,
meaning that (a) any control volume $\ibold^l\in\Omega^\ell$ is either
completely covered by $(n_{ref}^\ell)^{\bf D}$ finer control volumes or
by none, and (b) for any given pair of meshes $\Omega^{\ell-1}$ and
$\Omega^{\ell+1}$ there is always a layer of cells of $\Omega^{\ell}$
separating the two.  By analogy with the single grid case we can
construct the set $\Omega_f^{\ell,\edbold}$ corresponding to the faces in
$\Omega^\ell$, and likewise the set $\Omega_e^{\ell,\edbold}$ for the
corresponding edges. Similarly, for each level, we can define a 
divergence and curl operators for face and edge centered vectors, respectively.

The part of an AMR level which is ``covered'' by
refinement is denoted as the {\it covered region}, while the {\it
  valid region} of a given level $\ell$ is the part of $\Omega^\ell$
not covered by refinement. For computational convenience, solution
values are maintained in covered regions as well as valid regions, but
only the solution in valid regions is considered to be valid. The {\it
  composite solution} spans the computational domain and is the union
of the valid-region solutions on each level. The coarse-fine interface
between levels $\ell$ and $\ell-1$ is denoted by $\partial
\Omega^\ell$.

As in \cite{bergercolella89}, 
we refine in time as well as space, with $\Delta
t^{\ell+1} = \Delta t^{\ell} / n^{\ell}_{ref} $. The update of the
solution on the hierarchy of AMR levels can be described recursively as the
update of a single AMR level $\ell$ from time $t^\ell$ to time $t^\ell +
\Delta t^\ell$ (Fig.~\ref{fig:amrTimestep}).   
\begin{figure}
\begin{tabbing}
123\=456\=789\=012\=\kill
LevelAdvance($\ell, t^\ell, \Delta t^\ell$) \\
\> SingleLevelUpdate($\ell, t^\ell, \Delta t^\ell$): \\
\> \>  Interpolate solution values as needed from next coarser level  $(\ell-1)$  \\
\> \> Update solution on level $\ell$ using scheme described in Section
\ref{algo:se} \\
\> \> {\bf if}  ($\ell < \ell_{{\rm{max}}}$): increment fine flux registers \\
\> \> {\bf if}  ($\ell > 0$): increment coarser-level flux registers \\
\> \> $t^\ell := t^\ell + \Delta t^\ell$ \\
\> Recursively update any finer levels: \\
\> \> \bf{if} ($\ell < \ell_{{\rm{max}}}$) \bf{then} \\
\> \> \> $\Delta t^{\ell+1} = \frac{\Delta t^\ell}{n_{ref}^\ell}$ \\
\> \> \> { \bf{for} } $n = 1, n_{ref}^{ell}$ \\
\> \> \> \> LevelAdvance($\ell+1, t^{\ell+1}, \Delta t^{\ell+1}$) \\
\> \> \> { \bf {end for} } \\
\> \> \> ``Synchronize'' levels $\ell$ and level $\ell+1$ \\
\> \> { \bf{ end if} } \\
{\bf end} LevelAdvance
\end{tabbing}
\label{fig:amrTimestep}
\caption{Recursive AMR timestep}
\end{figure}

Extending the CT scheme described in this paper requires
some additions to the standard set of algorithmic tools generally used for
fully cell-centered discretizations of hyperbolic conservation laws like that
in \cite{bergercolella89}. 
Most of the additional algorithmic pieces 
result from the addition of the solenoidal face-centered $\vec{B}$
field. 

\subsection{Filling Ghost cells}
Before each single level update from time $t^\ell$ to time $t^\ell+\Delta t^\ell$, a ring of ``ghost'' cells sufficiently large to complete the
  stencils required to update valid-region data on each grid-patch is filled
  in. For the scheme described here, we require 6 and 4
  ghost cells for PPM and PLM reconstruction methods, respectively. Where possible, ghost values are filled by copying valid-region
  data from other grids on the same level $\ell$ or possibly by a discrete representation
  of physical domain boundary conditions. Where neither of these is possible,
  values must be interpolated from the coarser level $\ell-1$ solution. 
Cell-centered quantities are interpolated
  using a limited piecewise-linear scheme. The face-centered $\vec{B}$ field
  is likewise interpolated using a piecewise-linear scheme as follows:
\begin{enumerate} 
\item First, $\vec{B}^{\ell-1}$ is linearly interpolated in time to
  $t^\ell$. (Recall that level $\ell-1$ has already been updated from
  $t^{\ell-1}$ to $(t^{\ell-1} + \Delta t^{\ell-1})$, and $t^{\ell-1} \leq
  t^\ell < (t^{\ell-1} + \Delta t^{\ell-1})$. 
\item Then, fine-level
  faces which overlie coarse-mesh faces are filled using piecewise linear
  interpolation of co-planar coarse-mesh faces. 
\item Finally, values for fine-level faces which do not overlie coarse-mesh
  faces 
  are linearly interpolated between the two surrounding co-directional faces
  for which there 
  are already values (either fine-level faces on the coarse-fine boundary or
  fine-level values interpolated in step (1) ).
\end{enumerate} 
Note that we have not found it necessary to use a divergence-free
interpolation scheme like that described in \cite{Balsara2001JCoPh.174..614B}
to fill in values for ghost faces.  

\subsection{Synchronization}
After the sub-cycled advance of level $\ell+1$, the solutions on levels $\ell$
and $\ell+1$ have reached the same solution time $(t^\ell = t^{\ell+1})$, and
are then {\it synchronized}. For cell-centered conserved variables,
synchronization is identical to that used in \cite{bergercolella89}:
\begin{enumerate}
\item Replace level $\ell$ solution with the averaged level $\ell+1$ solution
  in covered regions. 
\item Because the fluxes used to update the fine-level solution were computed
  independently of those used to compute coarse-level updates, conservation
  will not be maintained at coarse-fine interfaces. In \cite{bergercolella89}
  and \cite{martincolella00}, {\it flux registers} are defined along the
  coarse-fine interface between levels $\ell$ and $\ell+1$, in which the
  fluxes used to compute coarse- and fine-level updates are stored. Since we
  consider the fine-level fluxes to be more accurate, we update the solution
  in the coarse cells adjoining the coarse-fine interface with the
  ``reflux-divergence'' of the difference of the fluxes:
\begin{equation}
U^\ell := U^\ell - \Delta t^\ell D_R^\ell(\sum_{n=1}^{n_{ref}^\ell} \langle
F^{\ell+1} \rangle - F^\ell) 
\end{equation}
where $U^\ell$ is the vector of conserved quantities, $D_R$ is the ``reflux
divergence'' operator, $F^\ell$ is the vector of fluxes used to update
$U^\ell$, $\langle F^{\ell+1} \rangle$ is the average of the level $(\ell +
1)$ fluxes on 
the underlying level $\ell$ faces, and the sum is over sub-cycled $\ell+1$
timesteps. 
\end{enumerate}

Synchronization for the face-centered magnetic field looks similar, and takes
the same form as described in \cite{Balsara2001JCoPh.174..614B}:
\begin{enumerate}
\item Replace level $\ell$ magnetic field $\vec{B}^\ell$ with the averaged
  level $\ell+1$ solution on covered faces.
\item Because the edge-centered electric fields on each level are computed
  independently, the composite $\vec{B}$ field will no longer be
  solenoidal. We treat this using an analogue to the face-centered flux
  registers used for cell-centered data, as presented in
  \cite{Balsara2001JCoPh.174..614B}. We store the  edge-centered electric
  fields along coarse-fine interfaces, then increment the coarse-level
  magnetic field at faces bordering the coarse-fine interface with a {\it
    reflux-curl} operator applied to the coarse-fine mismatch in the
  edge-centered electric fields. 
\end{enumerate}

\subsection{Regridding}

It is often desirable for refined regions to periodically adapt as the
solution evolves in time. When newly refined regions are created,
cell-centered fields are also interpolated from coarse-level data
using limited piecewise-linear interpolation. For $\vec{B}$ field
values of newly refined faces, we use the divergence-free
interpolation scheme described
in~\cite{Balsara2001JCoPh.174..614B}. The scheme is defined for
refinement ratios $n_{ref}=2$, but it can be applied recursively for
larger values of the refinement ratio.

\subsection{Shock-Cloud Interaction}

As an example of an AMR-MHD application we compute the interaction of a
cloud with a strong shock wave. This process is common in the
interstellar medium where shocks produced by supernova explosions
interact with the surrounding multi-phase
medium~\citep{klein1994ApJ...420..213K,MacLow1994ApJ...433..757M}.
Related processes, characterized by similar hydrodynamic structures,
are the supersonic motion of an over-dense cloud through a thin
magnetized
medium~\citep{Schiano1995ApJ...439..237S,Jones1996ApJ...473..365J,Vietri1997ApJ...483..262V,mrfj99,gmrj99,gmrj00},
or supersonic clouds
collisions~\citep{Lattanzio1985MNRAS.215..125L,Klein1995ASPC...80..366K,mjfr97,mijory99}.

The initial conditions for our problem are as follows: the background
gas has unit density and thermal pressure, and is at rest; a cloud
with the same pressure but 10 times higher density, moves through the
thin gas with a velocity $v_c=-3.47871373$ along the x direction.  A
plane-parallel shock with Mach number ${\cal M}=10$ propagates along
the same axis but in the opposite direction to the cloud.  The initial magnetic field is
uniform, of unit strength and aligned with the $x-$axis. The
computational box has dimensions $[0,1]\times[0,0.5]\times[0,0.5]$.
The boundary conditions correspond to supersonic inflow and outflow
for the lower and upper boundaries of the $x-$axis, and are periodic
otherwise.  The calculation is carried out in three dimensions on a base
grid of $64\times 32\times 32$ cells. Two additional levels of
refinement, with refinement ratio 4, are generated dynamically in
regions where the normalized undivided density 
gradient $|\Delta\rho|/\rho >0.1$,
and/or in the presence of shocks according to the criteria $|\Delta
P|/P >0.1$ and $\nabla\cdot {\bf v}<0$.

\begin{figure*}
\plotone{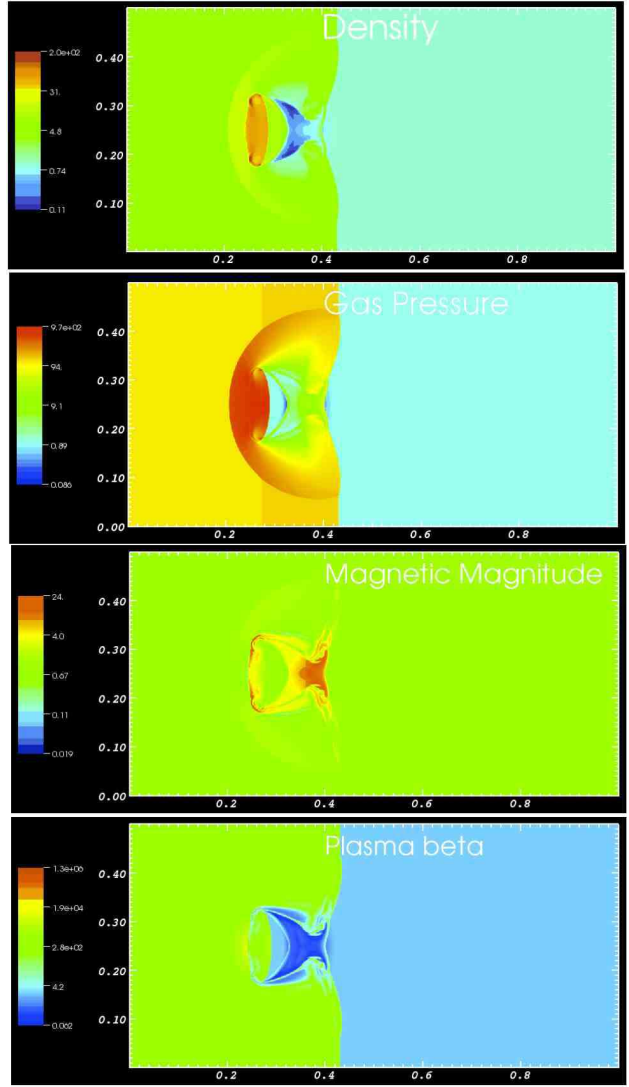} \\
\caption{Shock-cloud interaction: from top to bottom, solution for
density, gas pressure, magnetic field magnitude and plasma beta
parameter, at time $t=0.021251$.}
\label{cloud:fig}
\end{figure*}

Fig.~\ref{cloud:fig} shows from top to bottom the solution for the
density, gas pressure, magnetic field magnitude and plasma beta
parameter ($\beta=P_{gas}/P_B$), at time $t=0.021251$, corresponding to
160 cycles on the finest level. The main features, discussed at length
in the above papers, are correctly reproduced. The plane shock
front moving from the left crushes the cloud. As the cloud moves to
the left, it creates a bow shock in front of it, where the pressure
has its highest value. The cloud motion also generates a low pressure
region at its rear, where the magnetic pressure dominates the gas
pressure and the beta plasma is much less then unity. In addition, as
the fluid flows past the cloud, the magnetic field lines entrained in
the cloud body fold on themselves creating a current
sheet~\citep{Jones1996ApJ...473..365J}. 

Note that the maximum value of
the normalized divergence of the magnetic field $|\Delta x\nabla\cdot{\bf
  B}|/|B|$, is a few$\times 10^{-13}$. This is completely negligible
with respect to the solution value and demonstrates that our
implementation of the above operators for the coarse fine magnetic
field interpolation and refluxing operations is correct.

\section{Extension to Cosmology} \label{cosmol::se}

We now describe the extension of our MHD algorithm to the case of cosmological
applications. This will include only a basic description of the cosmological
code, for it is presented in detail elsewhere~\citep{mico07b}.

For cosmological simulations it is preferable to transform away the
expansion of the universe through the use of a {\it comoving} frame of
reference.  Thus we operate the transformation
\begin{equation}\label{comfr.eq}
{\boldsymbol x} \leftarrow a(t)^{-1}\;{\boldsymbol x} 
\end{equation}
from lab to comoving coordinates, where $a(t)$ is a function of time
that defines the physical size of spatial scales, and $\dot{a}/a$ is
the expansion rate of the universe. In addition we subtract out the
velocity component due to the expansion of the universe, and retain the
{\it peculiar} proper velocity, i.e.
\begin{equation}
{\boldsymbol u} \leftarrow
{\boldsymbol u} - \dot{a} \,{\boldsymbol x}.
\label{pecvel:eq}
\end{equation}
Finally, it is convenient to use {\it comoving} density and pressure,
i.e. those expressed in terms of the comoving volume ${\boldsymbol
  x}^3$ as opposed to the proper volume $a^3 {\boldsymbol x}^3$,
\begin{eqnarray}
\label{procom1.eq}
\rho &\leftarrow a^3 \,\rho,\\
   p &\leftarrow a^3 \, p.
\label{procom2.eq}
\end{eqnarray}
Similarly, although the magnetic flux naturally scales with $a^2(t)$, we
transform it to a pseudo-comoving variable
\begin{equation} \label{ba.eq}
B \leftarrow a^\frac{3}{2} \,B.
\end{equation}
The above transformations allow writing
the conservation and induction equations in a form that, except for
the appearance of source terms, closely resembles the
original ones. This similarity allows us not only to solve for the MHD
system of equations in the cosmological framework with the same
algorithm described in Sec.~\ref{numsche::se} but also to apply the
same menagerie of AMR tools described in Sec.~\ref{amr::se} with
virtually no modification.

In fact, in the comoving frame, ${\boldsymbol x}$, the conservation equations
read
\begin{equation}\label{cosm_cons_sys:eq}
\frac{\partial U}{\partial t} + \frac{1}{a(t)}{\bf\nabla_x\cdot F} = S,
\end{equation}
where $U$ and $F$ are defined exactly as in (\ref{cons_uf:eq}) but are
now expressed in terms of peculiar velocity, comoving density, comoving
pressure and pseudo-comoving magnetic field. The source term on the RHS is 
\begin{equation}
S(U) = -\,\frac{\dot a}{a}
\begin{pmatrix}
0 \\
\rho u_{0} \\
\vdots \\
\rho u_{\Dim-1} \\
2\rho e-\frac{B^2}{2} \\
\half B_0 \\
\vdots \\
\half B_{\Dim-1}
\end{pmatrix}
+\rho
\begin{pmatrix}
0 \\
g_0 \\
\vdots \\
g_{\Dim-1} \\
{\bf u\cdot g} \\
0 \\
\vdots \\
0
\end{pmatrix},
\label{eqn:source}
\end{equation}
where the first term, $\propto\dot a/a$, accounts for adiabatic losses
of momentum, energy, and magnetic field, and the second term,
${\propto\bf g}$, is due to gravity.  Similarly, we rewrite
Faraday's law in the comoving frame in terms of
the peculiar velocity and pseudo-comoving magnetic field, i.e.
\begin{equation} \label{cosm_faraday:eq}
\frac{\partial B_d}{\partial t} 
= -\frac{1}{a} 
\frac{\partial}{\partial x_j} \left(u_jB_d-B_ju_d\right)
-\frac{1}{2}\frac{\dot a}{a} {B_d}.
\end{equation}
Based on Eq. (\ref{cosm_cons_sys:eq}) and (\ref{cosm_faraday:eq})
the time update of $U$ and $B$ is then done according to
\begin{eqnarray}
U^{n+1}_\ibold&=& U^n_\ibold -\frac{1}{a^{n+\half}}\frac{\Delta t}{\Delta x}
\left(\nabla\cdot F\right)_\ibold^{n+\half} + \Delta t \,S_\ibold^{n+\half}, \\
 B^{n+1}_{d,\ibold +\half \ebold^d}&=& \left(\frac{a^n}{a^{n+1}}\right)^\frac{1}{2}
 B^n_{d,\ibold +\half \ebold^d} -\frac{1}{(a^{n+\half}a^{n+1})^\half}\frac{\Delta t}{\Delta x}
\left(\boldsymbol D\times \boldsymbol E\right)_{d,\ibold +\half \ebold^d}^{n+\half},
\end{eqnarray}
where the time centered fluxes and electric fields, as well as the synchronization between face and cell centered magnetic field variables, are computed using
the algorithm defined in Sec.~\ref{numsche::se}. A second order estimate of
the source term can be obtained by the simple time average
$S^{n+\half}\simeq\half(S^n+S^{n+1})$.  In reality, the source terms
associated with gravity and expansion are implemented using a slightly
more sophisticated method that estimates the change in kinetic energy
due to the work by gravity, directly from the change in the momentum
components. This is described in detail in~\cite{mico07b}. Similarly,
after the face-centered magnetic field variables have been updated in
time, the cell-centered values are synchronized, and the change in
magnetic field energy due to cosmic expansion is computed from the
corresponding change in the magnetic field components.

\subsection{MHD Santa Barbara Test}

In this final test we present an MHD version of the `Santa Barbara
Cluster Comparison Project'. The tests consists of the formation of a
massive cluster of galaxies in a 64 Mpc volume. The cosmological model
is an Einstein-De Sitter universe (i.e. with critical total matter density)
with 10\% of the total matter density in baryons, and an expansion
rate given by a Hubble parameter, $H_0 = 50$ km s$^{-1}$ Mpc$^{-1}$
(see additional details in~\cite{frenketal99}).
The purpose of this calculation is to test our MHD solver 
in a realistic cosmological application. To compare 
with previously published results, the dynamic role of the magnetic field
remains negligible throughout the calculation, which we ensure by
adopting a sufficiently small initial magnetic seed. The geometry of such 
fields is immaterial, and is chosen to be uniform for convenience.

The calculation is performed basically as in~\cite{mico07b}, except
for the initial redshift which is $z = 30$ (instead of 40).  So, two
initial grids are in place at simulation start: a base grid covering
the entire 64 Mpc$^3$ domain with 64$^3$ cells and 64$^3$ particles;
and a second grid, also with 64$^3$ cells and 64$^3$ particles, but
only 32 Mpc on a side and placed in the central region of the base
grid, thus yielding an initial cell size of 0.5 Mpc.  Refinement is
applied only within the latter higher resolution region to cells with
a total mass larger than $6.4\times 10^{10}$ M$_\odot$.  We allowed
for 5 additional levels of refinement (for a total 6 level hierarchy), with a
constant refinement ratio $n_{ref}=2$.  The size of the finest mesh is
about 15 comoving kpc.  The timestep is limited by the most stringent
among the following three conditions: the Courant-Friedrichs-Lewy
condition on the MHD waves, with coefficient $C_{\rm CFL}=0.8$, an
analogous {\rm CFL} condition based on the speed of the collision-less
particles, with coefficient $C_{\rm part} =0.5$, and the requirement
that the expansion of the universe during a time-step is less than
2\%. The calculation was performed using the PPM reconstruction scheme
and the HLLD Riemann solver.

\begin{figure*}
\plotone{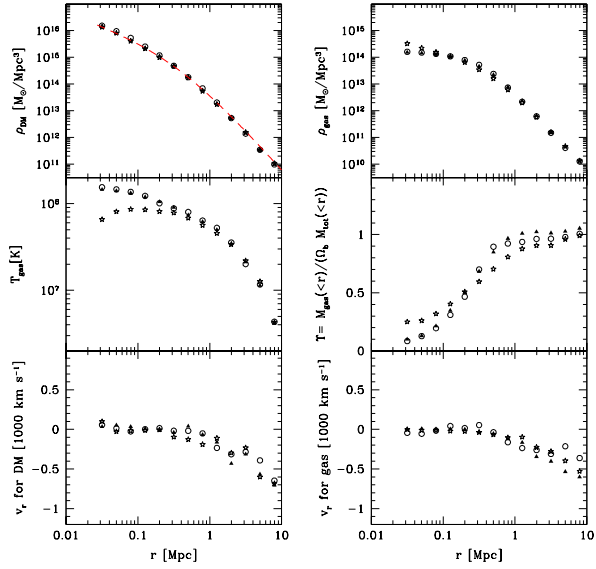} \\
\caption{Radial profile of dark matter (top left), baryonic gas (top
  right) temperature (middle left), baryonic fraction (middle right),
  radial velocity for dark matter (bottom left) and gas (bottom
  right).  In addition to the results from {\tt CHARM} (open circles),
  for comparison we also show those from the {\tt ENZO} AMR code
  (filled triangles)~\cite{bryannorman97} as well as those from the
  {\tt HYDRA} SPH code (open stars)~\cite{cope95}. }
\label{sb:fig}
\end{figure*}

\begin{figure*}
\epsscale{1}\plottwo{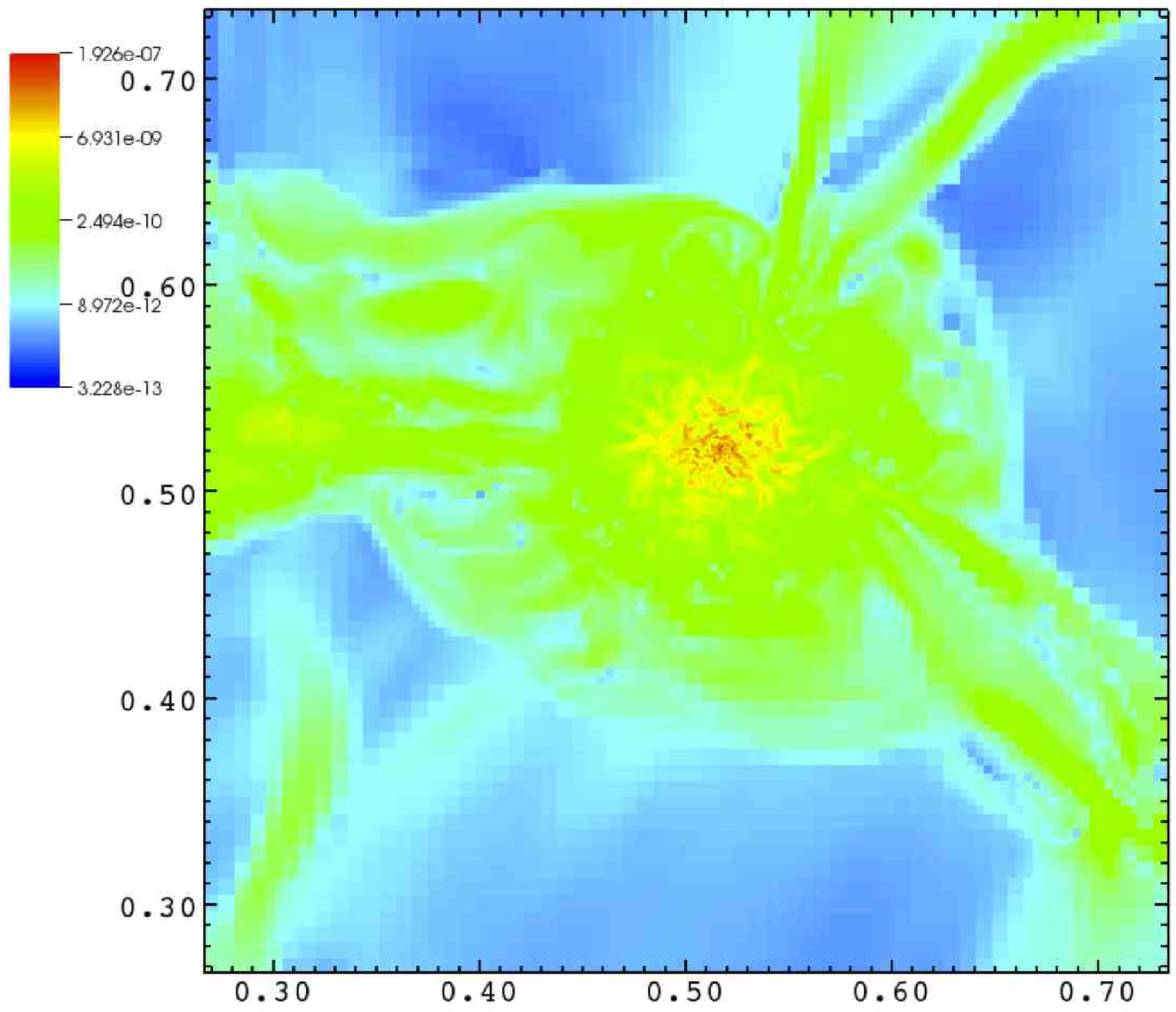}{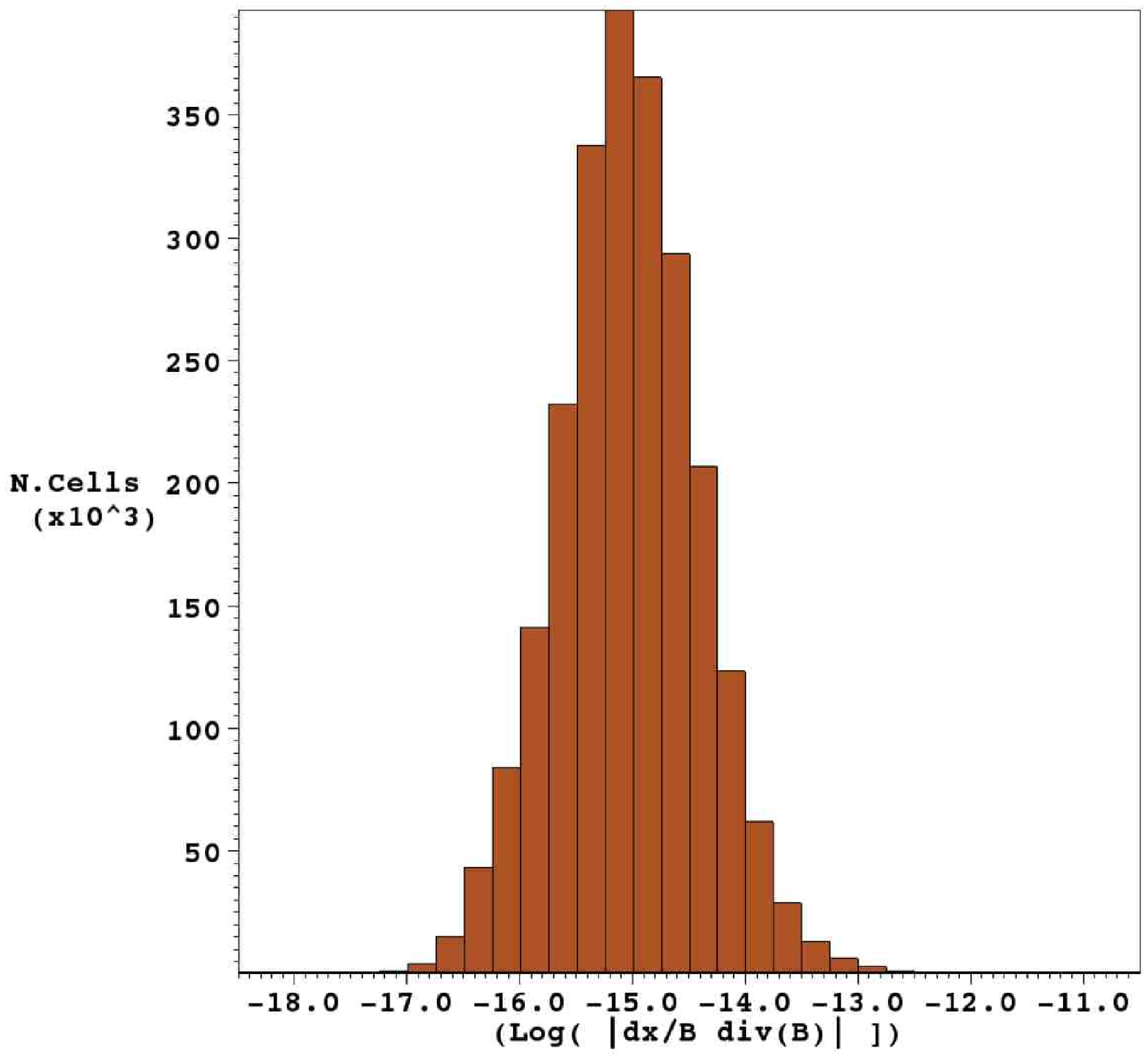} \\
\caption{Left: Distribution of the magnetic field magnitude on a plane across the
center of the simulated cluster. Right: Histogram of the absolute value of the normalized
divergence of the magnetic field, $|(\Delta x/B) \nabla B|$.}
\label{sb2:fig}
\end{figure*}

The results of the calculation are summarized by the radial plots in
Fig.~\ref{sb:fig}, where in analogy to~\cite{mico07b} we show results
from two other simulation codes.  As expected, the MHD solver (in the
limit of vanishing magnetic field) produces virtually the same results as the 
hydro solver in~\cite{mico07b}, attesting therefore to the same
reliability for highly nonlinear calculations,
involving high Mach number flows and large dynamic range in the fluid
quantities.
Finally, the left panel of Fig.~\ref{sb2:fig} shows the magnetic field
magnitude (in arbitrary units) distribution on a slice passing through
the cluster center.  The magnetic field is stronger in the core region
where it also shows substantial spacial structure down to the smallest
resolvable scales.  On the left panel of the same Figure we present
the distribution of the normalized magnetic field divergence,
$|(\Delta x/B) \nabla B|$.  The bulk of the distribution is at the
level of $10^{-15}$ or so, and the max value is around
$10^{-11}$. Again, this value is completely negligible with respect to
the solution value.  While larger than the value obtained in the
previous test example, this is expected given the much larger number of
integration steps (about 10$^4$) in the current case.

\section{Summary}\label{summary:se}

We have presented the implementation of a three-dimensional scheme for
MHD in the AMR code {\tt CHARM}.  The scheme uses a hybrid
discretization, in the sense that fluid quantities are cell-centered,
and magnetic field variables are face-centered. 
The algorithm is based on the full 12-solve spatially unsplit
Corner-Transport-Upwind (CTU) scheme~\citep{Colella1990JCoPh..87..171C}. 
The fluid quantities are updated using the PPM method, while the magnetic
field evolution is computed through a CT method.  The edge-centered electric fields
necessary for the CT step are computed as in~\cite{GardinerStone2005JCoPh.205..509G}.
We employ a
simplified version of the multidimensional MHD source terms required
in the predictor step for high-order 
accuracy~\citep{GardinerStone2005JCoPh.205..509G,GardinerStone2008JCoPh.227.4123G}, 
which is as in~\citet{Crockett2005JCoPh.203..422C}. This greatly
simplifies the three-dimensional version of the algorithm with respect
to the original form, without compromising the accuracy and robustness
of the solutions.

The algorithm is implemented in an AMR framework. This requires
synchronization operations across refinement levels, including
face-centered restriction and prolongation operations and a reflux-curl
operation, which is necessary to maintain a divergence-free magnetic
field solution~\cite{Balsara2001JCoPh.174..614B}. 
The code works with any even refinement ratio, although
values above 4 are unusual. Our tests demonstrate that the code
converges at the expected rate, is robust in problems involving strong
shocks, maintains the magnetic field divergence at a negligible value
and is suitable for astrophysical and cosmological applications.



\bibliographystyle{apj}
\bibliography{../biblio/books,../biblio/codes,../biblio/papers,../biblio/proceed}

\end{document}